\documentclass[a4paper,11pt]{article}
\pdfoutput=1 

\usepackage{jcappub} 

\usepackage[T1]{fontenc} 

\usepackage{latexsym}    
\usepackage{graphicx}   
\usepackage{verbatim}   
\usepackage[normalem]{ulem} 
\usepackage[dvipsnames]{xcolor}     
\usepackage{tikz}       
\usetikzlibrary{shapes}
\usepackage{subfigure}  
\usepackage{multirow}
\usepackage{hhline}
\usepackage{feynmp-auto}
\usepackage{url}
\usepackage{hyperref}   
\usepackage{tensor}     
\usepackage{amssymb}
\usepackage{enumerate}	
\usepackage{braket}		
\usepackage{physics}
\usepackage{color}

\newcommand{\Rev}[1]{{#1}}
\newcommand{\COMRev}[1]{}  

\title{\boldmath Observational constraints on generalised axion-like potentials for the late Universe}

\author[a]{Hsu-Wen Chiang,}
\author[b]{Carlos G. Boiza,} 
\author[c,b]{Mariam Bouhmadi-López}

\affiliation[a]{Department of Physics, Southern University of Science and Technology, Shenzhen 518055, China}
\affiliation[b]{Department of Physics \& EHU Quantum Center, University of the Basque Country UPV/EHU, P.O. Box 644, 48080 Bilbao, Spain}
\affiliation[c]{IKERBASQUE, Basque Foundation for Science, 48011, Bilbao, Spain}

\emailAdd{jiangxw[at]sustech.edu.cn, b98202036[at]ntu.edu.tw}
\emailAdd{carlos.garciab[at]ehu.eus}
\emailAdd{mariam.bouhmadi[at]ehu.eus}

\abstract{Axions have emerged as compelling candidates for describing the dark sector of the Universe. In this work, we explore quintessence models inspired by axion-like potentials as a dynamical alternative to the cosmological constant. These models naturally exhibit a tracking behaviour, reducing the need for fine-tuned initial conditions.
We perform a Markov chain Monte Carlo (MCMC) analysis on a complete cosmological dataset of Planck PR4 cosmic microwave background (CMB), DESI DR1 baryon acoustic oscillation (BAO), Pantheon+ type Ia supernovae, low-$z$ Cepheid anchors, and DES Y1 large scale structure measurement. Unfortunately, neither the Hubble tension nor the $S_8$ tension is eased. The model does reach parity with $\Lambda$CDM model statistically according to DIC, WAIC and Bayesian ratio, suggesting that a quintessence model may still help with the cosmic tension by further extending the model we have investigated. 
}

\begin{document}
\maketitle
\flushbottom

\section{Introduction}
\label{sec:intro}


We are currently living an outstanding time in the field of cosmology
and gravitation where more and more data are being collected constantly. One of
the most puzzling questions that we cosmologists are facing nowadays is the
mysterious source that is driving our recent acceleration. In fact, 
the acceleration was first inferred from measurements of Type Ia supernovae almost 30 years ago \cite{SupernovaSearchTeam:1998fmf,SupernovaCosmologyProject:1998vns,Bahcall:1999xn} and corroborated later on through measurements from the cosmic microwave background (CMB) \cite{Bennett2003,Komatsu2011,Planck2013,Planck2018},  baryon acoustic oscillations (BAO) \cite{Eisenstein2005,Cole2005,Percival2010} and large-scale structure observations \cite{Tegmark2004,Reid2010}.

The most straightforward explanation for the late-universe accelerating expansion is the presence of dark energy—a mysterious form of energy that pervades space and exerts negative pressure. At low redshift, this negative pressure drives the accelerated stretching of space-time. Dark energy is commonly modelled as a positive cosmological constant $\Lambda$, representing a uniform and constant energy density that permeates all of space and exerts negative pressure everywhere. In addition, plenty of extended theories of gravity can easily accommodate a late-time acceleration of the Universe through some effective dark energy fluid (cf. ref.\cite{CANTATA:2021asi} for a review on this topic). Regardless of the approach used to describe the current acceleration of the Universe, we know that the substance driving this accelerated expansion now constitutes about two-thirds of the total cosmic content. The remaining third is primarily dark matter (DM), with a smaller fraction in the form of baryonic matter.


This standard cosmological model ($\Lambda$CDM) remains the most widely accepted framework to describe the late-time acceleration of the Universe. However, its dominance has been challenged by tensions in measurements of key cosmological parameters, as observations from different datasets yield conflicting results. These discrepancies may point to new physics beyond the $\Lambda$CDM model or systematic errors in the measurements \cite{DiValentino2021,Abdalla:2022yfr}. Two of the most prominent cosmological tensions are the Hubble tension and the $S_8$ tension\footnote{The parameter $S_8$ quantifies the clustering of matter in the Universe. It is defined as $S_8=\sigma_8\sqrt{\Omega_{m0}/0.3}$  where $\sigma_8$ is the amplitude of matter fluctuations on scales of 8 Mpc/h, and $\Omega_{m0}$ is the matter density at the present time.}.

The Hubble tension refers to the discrepancy between different measurements of the Hubble constant, $H_0$, which quantifies the Universe's expansion rate. Specifically, early-universe indirect measurements inferred from the CMB acoustic scale yield a lower value of $H_0 = 67.26 \pm 0.49$ km/s within the framework of flat $\Lambda$CDM model \cite{Rosenberg:2022sdy} than the higher value of $H_0 = 73.04 \pm 1.04$ km/s directly derived from Type Ia supernovae (SNe) light curve and the low-$z$ SNe light curve anchors such as Cepheid variables \cite{Riess:2020fzl, Brout:2022vxf}, leading to a significant $5.0\sigma$ tension between $H_0$ values derived from these two observations.

On the other hand, the $S_8$ tension refers to the discrepancy between values of the rescaled matter perturbation strength $S_8$, with early Universe estimates of $S_8 = 0.827 \pm 0.013$ from the CMB power spectrum yielding higher values than late-time measurements of $S_8 = 0.776 \pm 0.017$ from weak gravitational lensing surveys and galaxy clustering \cite{DES:2021wwk}, leading to a statistically relevant $2.4\sigma$ tension between two completely orthogonal datasets.

Coincidentally or not, both $H_0$ and $S_8$ tension are sourced by the early-time estimates based on CMB physics and a presumed cosmic evolution, more specifically the cosmological standard model of $\Lambda$CDM, and the late-time counterparts that does not rely on assumption about the cosmic evolution. This separation of time scale makes it a perfect testing ground for either early-time modification to CMB physics, or late-time modification to the cosmic evolution.


\COMRev{Besides}\Rev{The observational challenges and opportunities call for theoretical investigation. In particular}, axion-inspired scalar field models have been widely explored in cosmology as candidates for both dark energy and dark matter, with applications spanning from early universe inflation to present-day cosmic acceleration. These models often involve potentials motivated by axion physics, which naturally arise in several fundamental theories. Among them, a widely studied class of models adopts potentials of the form $V(\phi)\propto\left[1-\cos{(\phi/\eta)}\right]^{-n}$. Such potentials emerge in high-energy frameworks and have been investigated in various cosmological scenarios. The case where $n<0$ has been extensively explored in the literature, including applications in the string axiverse as a solution to the coincidence problem \cite{Kamionkowski:2014zda,Emami:2016mrt}, in early dark energy models as a means to address the Hubble tension \cite{Poulin:2018cxd,Kamionkowski:2022pkx}, in Natural Inflation models \cite{Freese:1990rb}, and in primordial black hole formation mechanisms \cite{Khlopov:2004sc}. Additionally, the same potential has been used in axionic dark matter studies \cite{Khlopov:1999tm,Hui:2016ltb}, where oscillations around the potential minimum play a crucial role in defining the dynamics of ultralight axions. Cosmological simulations involving such ultralight axions have been performed in \cite{Schive:2014dra,Schive:2014hza,Mocz:2019uyd}, demonstrating their viability as dark matter candidates. For further discussions on axions in cosmology, see \cite{Ng:2000di,Hammer:2020dqp,Marsh:2015xka,Chakraborty:2021vcr}.


In this work, we focus on a specific dark energy model within this broad axion-inspired framework: the generalised axion-like dark energy model, first introduced in \cite{Hossain:2023lxs} and further analysed in \cite{Boiza:2024azh,Boiza:2024fmr,Hossain:2025grx}. In contrast to the more commonly studied $n<0$ case, here we analyse the scenario where $n>0$. The terminology generalised axion-like stems from its resemblance to the axion potentials appearing in particle physics, with the key distinction that we consider a positive $n$. The studies presented in \cite{Hossain:2023lxs,Boiza:2024azh} demonstrated that this potential exhibits tracking behaviour, making it a compelling alternative to other quintessence models. Tracking solutions allow the scalar field to dynamically adjust to the evolution of the dominant energy component in the universe, alleviating the coincidence problem. Additionally, this model provides a \COMRev{natural }mechanism for a smoother transition from matter domination to dark energy, making it preferable over alternative quintessence potentials. Further comparisons with the inverse power-law potential were made in \cite{Hossain:2023lxs}, where it was shown that the generalised axion-like potential leads to a more rapid transition from matter to dark energy while ensuring that the present-day equation of state remains closer to $w=-1$. These properties suggest that this model provides a promising alternative to both $\Lambda$CDM and traditional quintessence models\Rev{, by substituting the cosmological constant with a dynamical field that is free from additional fine-tuning}.


The goal of this work is to put the axion-like dark energy model to test, in particular how useful it is for releasing the current cosmological tensions. For this purpose, we conduct a comprehensive Markov Chain Monte Carlo (MCMC) analysis on cosmological datasets from various sources, including CMB (Planck2018 low-$\ell$ TTEE \cite{Planck:2019nip}, NPIPE CamSpec high-$\ell$ TTTEEE \cite{Rosenberg:2022sdy}, PR4 CMB lensing \cite{Carron:2022eyg, Carron:2022eum}) , BAO (DESI DR1 without full-shape \cite{DESI:2024mwx, DESI:2024lzq, DESI:2024uvr}), supernovae (Pantheon+ SNe catalog \cite{Brout:2022vxf} without anchoring of SNe standardised absolute magnitude), low-$z$ $H_0$ measurements (Riess et al. low-$z$ anchors \cite{Riess:2020fzl}), and DES Y1 morphology \cite{DES:2017myr}. While prior works \cite{Hossain:2023lxs, Hossain:2025grx} on the cosmological MCMC analysis of axion-like dark energy model do exist, the choice of parameter range and the incompleteness of the MCMC analysis prompts us to revisit this issue.

To verify if the tension is reduced, the model prediction of the relevant parameters such as $H_0$, $\Omega_{m0}$ and $S_8$ from the axion-line dark energy model is checked against the prediction of $\Lambda$CDM model. Parameters irrelevant to the cosmological tension are checked as well to avoid any adverse effect. Furthermore, we introduce statistical probes recommended in prior researches \cite{DES:2020hen, Raveri:2019gdp, PhysRevD.100.023512} such as Bayesian ratio, suspiciousness, and goodness of fit as alternatives to tension analysis.
Finally, we utilise CAMB Boltzmann solver \cite{Lewis:1999bs,Howlett:2012mh} to generate CMB power spectrum, $f\sigma_8$ and matter power spectrum. The results are compared with observations \cite{eBOSS:2020lta, Turnbull:2011ty, Achitouv:2016mbn, Beutler:2012px, Feix:2015dla, BOSS:2016wmc, BOSS:2013eso, Blake:2012pj, Nadathur:2019mct, BOSS:2013mwe, Wilson:2016ggz, eBOSS:2018yfg, Okumura:2015lvp, Reid:2009xm, eBOSS:2018qyj} to tell if there are any effect on the large scale structure formation.
    
    
Our paper is structured as follows. In section~\ref{sec:model}, we introduce the model that will be observationally constrained. This model is based on a quintessence model with an axion-like potential, emphasising its tracking regime, a crucial feature for addressing the coincidence problem and mitigating the need for fine-tuned initial conditions. In section~\ref{sec:fit}, we perform the observational fit, detailing the methodology and datasets used, and analyse carefully how the tracking dark energy we consider affects the cosmological parameters, and thus the tensions. We also discuss the possible mechanism behind in depth. Finally, in section~\ref{sec:conclusion}, we summarise what we find about the axion-like dark energy model through our MCMC analysis.


\section{Generalised axion-like potentials}
\label{sec:model}

In this section, we present the generalised axion-like model, which we propose as a framework to explain the late-time accelerated expansion of the universe. Its dynamics and cosmological perturbations have been extensively analysed theoretically recently in \cite{Hossain:2023lxs,Boiza:2024azh,Boiza:2024fmr,Hossain:2025grx}. This model relies on a scalar field minimally coupled to gravity, classified as a quintessence model. It is particularly noteworthy for its ability to address the coincidence problem through tracking behaviour, as well as by favouring a rapid transition to a dark energy-dominated universe at late-time \cite{Boiza:2024azh}.

\subsection{Model description}

We begin by defining the action for our model, which combines contributions from gravity, matter, radiation, and the scalar field:

\begin{equation}\label{scalaraction}
   S=\int d^4x\sqrt{-g}\left(\frac{R}{2k^2}+L_m+L_r+L_{\phi}\right),
\end{equation}
where $k^2=8\pi G=1/M_P^2$ (we use natural units $\hbar=c=1$ throughout this work and $M_P\approx2.435\times10^{18}\,\textrm{GeV}$ is the reduced Planck mass). $R$ denotes the Ricci scalar, and $L_m$, $L_r$, and $L_{\phi}$ correspond to the Lagrangian\COMRev{s} \Rev{densities }for matter, radiation, and the scalar field, respectively. For the scalar field, we define the Lagrangian \Rev{density }as

\begin{equation}\label{canonicallagrangian}
   L_{\phi}=-\frac{1}{2}g^{\mu\nu}\partial_{\mu}\phi\partial_{\nu}\phi-V(\phi),
\end{equation}
with the potential $V(\phi)$ given by

\begin{equation}\label{axionscalarpotential}
   V(\phi)=\Lambda^4\left[1-\cos{(\phi/\eta)}\right]^{-n},
\end{equation}
where $\Lambda$, $\eta$, and $n$ are positive free parameters of the model. \Rev{In standard axion-like scenarios, $\eta$ is typically associated with the spontaneous symmetry-breaking scale of a global $U(1)$ symmetry, while $\Lambda$ represents the scale of non-perturbative effects that generate the potential. The exponent $n$ parametrises deviations from the usual cosine potential.}

To describe the universe’s dynamics, we adopt a spatially flat FLRW metric:

\begin{equation}\label{rwmetric}
   ds^2=-dt^2+a^2(t)d\bar{x}^2.
\end{equation}
Here, $a(t)$ represents the scale factor. The scalar field  evolves according to the equation of motion

\begin{equation}\label{sfeq}
   \ddot{\phi}+3H\dot{\phi}+V_{,\phi}=0,
\end{equation}
where $H$ is the Hubble parameter, and $V_{,\phi}$ is the derivative of $V$ with respect to $\phi$. A dot denotes a derivative with respect to the cosmic time $t$. The Friedmann equation, which includes contributions from all energy components, is expressed as

\begin{equation}\label{friedmanneq}
   H^2=\frac{k^2}{3}\left[\rho_m+\rho_r+\frac{1}{2}\dot{\phi}^2+V(\phi)\right].
\end{equation}
The last two terms of the equation can be recognised as the energy density of the scalar field. For the sake of completeness, we also define the pressure of the scalar field. They respectively read as

\begin{equation}\label{densityandpressure}
   \rho_{\phi}=\dot{\phi}^2/2+V(\phi), \quad p_{\phi}=\dot{\phi}^2/2-V(\phi).
\end{equation}

\Rev{The background evolution of this model can be fully described using a dynamical system approach. By introducing suitable compact variables, the equations of motion can be recast into a closed autonomous system. This allows us to identify critical points, study their stability, and characterise attractor solutions (see table \ref{table:fixedpointsmodel}). In particular, the model dynamics can be described by the following set of differential equations \cite{Boiza:2024azh}:}

\Rev{\begin{equation}\label{xfinaleq}
   x'=\frac{1}{2}\left[-3x+3x^3-3xy^2+xz^2\right]+\sqrt{\frac{3}{2}}\lambda y^2,
\end{equation}}

\Rev{\begin{equation}\label{yfinaleq}
   y'=\frac{1}{2}\left[3y-3y^3+3yx^2+yz^2\right]-\sqrt{\frac{3}{2}}\lambda xy,
\end{equation}}

\Rev{\begin{equation}\label{zfinaleq}
   z'=\frac{1}{2}\left[-z+z^3+3zx^2-3zy^2\right],
\end{equation}}

\Rev{\begin{equation}\label{lambdafinaleq}
   \lambda'=-\sqrt{6}f(\lambda)x,
\end{equation}
where a prime denotes a derivative with respect to $\ln(a/a_0)$, and $a_0$ is the current value of the scale factor. \Rev{ Note that this set of equations corresponds to a reformulation of the original Friedmann and Klein–Gordon equations in terms of dimensionless variables, and rewritten as a first-order autonomous system.} The variables $x$ and $y$ correspond, respectively, to the dimensionless kinetic and potential energy contributions of the scalar field \cite{Copeland:1997et}:}

\Rev{\begin{equation}\label{xyvariables}
   x=\frac{k\dot{\phi}}{\sqrt{6}H}, \quad y=\frac{k\sqrt{V}}{\sqrt{3}H}.
\end{equation}
We have also introduced the variable $z$, which describes the radiation component and is defined as}

\Rev{\begin{equation}\label{zvariable}
   z=\Omega_r^{1/2}=\frac{k\sqrt{\rho_r}}{\sqrt{3}H},
\end{equation}
and the parameter $\lambda$ \cite{Steinhardt:1999nw,Copeland:1997et,delaMacorra:1999ff,Ng:2001hs}:}

\Rev{\begin{equation}\label{lambdadef}
   \lambda=-\frac{V_{,\phi}}{kV},
\end{equation}
which for our potential becomes \cite{Boiza:2024azh}}

\Rev{\begin{equation}\label{lambdaphimodel}
   \lambda(\phi)=\frac{n}{k\eta}\frac{\sin{(\phi/\eta)}}{1-\cos{(\phi/\eta)}}.
\end{equation}
Finally, the function $f(\lambda)$ which appears in \eqref{lambdafinaleq} is given in our model by \cite{Boiza:2024azh}}

\Rev{\begin{equation}\label{flambda}
   f(\lambda)=\lambda^2[\Gamma(\lambda)-1]=\frac{n}{2k^2\eta^2}+\frac{\lambda^2}{2n},
\end{equation}
where the parameter $\Gamma$ represents the potential’s curvature and is defined as \cite{Steinhardt:1999nw,delaMacorra:1999ff,Ng:2001hs}}

\Rev{\begin{equation}\label{gammadef}
   \Gamma=\frac{VV_{,\phi\phi}}{V_{,\phi}^2},
\end{equation}
which for our potential becomes \cite{Boiza:2024azh}}

\Rev{\begin{equation}\label{gammamodel}
   \Gamma=1+\frac{1}{2n}+\frac{n}{2k^2\eta^2\lambda^2}.
\end{equation}}

\begin{table}[t]
    \centering
    \Rev{\begin{tabular}{||c|c|c|c|c|c|c||}
    \hline
       \:Point\: & \:$x$\: & \:\:$y$\:\: & \:$z$\: & $\lambda$ & $w_{\phi}$ & Stability\\
       \hline
       \hline
       A  & $0$ & $0$ & $1$ & \:$\lambda_{\star}$\: & \:Undefined\: & Saddle\\
       B  & $0$ & $0$ & $0$ & \:$\lambda_{\star}$\: & \:Undefined\: & Saddle\\
       C  & $0$ & $1$ & $0$ & $0$ & $-1$ & \:Stable if $n>0$\:\\
       D  & \:$x_{\star}$\: & $0$ & \:$z_{\star}$\: & \:Undefined\: & $1$ & Saddle if $n>0$\\
       \hline
    \end{tabular}}
    \Rev{\caption{\label{table:fixedpointsmodel}Fixed points of the system given by \eqref{xfinaleq}, \eqref{yfinaleq}, \eqref{zfinaleq} and \eqref{lambdafinaleq}. D is deduced by compactifying the $\lambda$ variable. In fact, $\lambda$ blows-up in this case (for more details, see \cite{Boiza:2024azh}). The quantities $x_{\star}$, $y_{\star}$ and $\lambda_{\star}$ are random values of the variables $x$, $y$ and $\lambda$ inside their domains which satisfy the equations of motion.}} 
\end{table}

\Rev{The system admits a set of fixed points whose properties provide insight into the qualitative dynamics of the model. These fixed points, summarised in table \ref{table:fixedpointsmodel}, correspond to distinct cosmological regimes: points A and B are associated with radiation and matter domination, respectively, and generally appear as saddle points in dynamical analyses of quintessence models. Point C is a potential-dominated fixed point corresponding to a de Sitter state and describes the late-time dark energy-dominated regime. Unlike A and B, the stability of point C depends on the form of the potential. In our model, it is stable for $n > 0$, and hence acts as the true attractor governing the asymptotic cosmic acceleration. Point D, by contrast, represents a kinetically dominated phase and arises in the regime where $\lambda \rightarrow \infty$. To consistently incorporate it into the phase space analysis, the variable $\lambda$ must be compactified (see \cite{Boiza:2024azh} for more details).}

\Rev{A key feature of our model is the tracking behaviour, which appears only if $n > 0$. This phase is associated with initial values $\lambda_i \gg 1$ (where the subscript $i$ denotes the initial condition hereafter), where the field slowly rolls down a steep potential slope and the curvature parameter $\Gamma$ remains nearly constant. In this regime, the system evolves along a family of \textit{instantaneous critical points} in the $(x, y, z, \lambda)$ phase space, solutions that are not true fixed points, but which act as quasi-attractors guiding the field evolution \cite{Ng:2001hs}. This behaviour can be sustained as long as $\Gamma > 1$ and $\Gamma \approx \text{const.}$ \cite{Steinhardt:1999nw}. Under these conditions, the scalar field equation of state remains below that of the dominant component. For instance, during matter domination ($w = 0$), it takes the approximate value \cite{Boiza:2024azh}}

\Rev{\begin{equation}\label{sfeosgammamodel}
   w_{\phi} \approx -\frac{1}{1+n}.
\end{equation}
This tracking mechanism allows the scalar field to evolve in a manner largely independent of its initial conditions.}

\Rev{As the universe evolves further, the scalar field transitions from the tracking phase to a state where it dominates the energy density. This transition occurs when $\lambda$ decreases and the field approaches the minimum of its potential. Near the minimum, the potential can be approximated as \cite{Boiza:2024azh}}

\Rev{\begin{equation}\label{minimumexpansion}
   V(\phi)\approx V_{\rm min}+\frac{1}{2}m^2(\phi-\pi\eta)^2,
\end{equation}
where we have defined $V_{\rm min}\equiv\Lambda^4/2^n$ and $m^2\equiv n\Lambda^4/(2^{n+1}\eta^2)$. Note that we can use $V_{\rm min}$ and $\Lambda$ interchangeably as a free parameter of the model. In the following section, we will use $V_{\rm min}$ in the study of the observational constraints. \Rev{On the other hand, the scalar mass $m$ directly controls the ending of the tracking phase. For the tracking to last for $O(H_0^{-1})$, we need $m^2/H_0^2 \sim n M_P^2 \eta^{-2} \sim O(1)$. The field strength $\phi$ thus potentially reaches super-Planckian scale. However, it does not immediately lead to trans-Planckian problem, as $\phi$ by itself is not an observable from the gravity point of view.} At the end of this stage, the kinetic energy term becomes negligible, and the scalar field behaves like dark energy, fuelling accelerated cosmic expansion. In terms of the dynamical system, this final regime corresponds to the approach toward a true critical point: the de Sitter attractor $(x, y, z, \lambda) = (0, 1, 0, 0)$, listed as point C in table \ref{table:fixedpointsmodel}. This point is stable for $n > 0$ and represents a late-time universe with $w_{\phi} = -1$. Unlike the tracking phase, where $w_{\phi}$ depends on $n$, the de Sitter attractor yields a behaviour indistinguishable from a cosmological constant. The rapid growth of $\Gamma$ as the field approaches the minimum ensures the system settles quickly into this final state, distinguishing the generalised axion-like model from simpler inverse power-law potentials \cite{Hossain:2023lxs,Boiza:2024azh}.}

\subsection{Cosmological evolution}

In this subsection, we present some results obtained from the numerical study of the dynamical system of the generalised axion-like model \Rev{given by \eqref{xfinaleq}, \eqref{yfinaleq}, \eqref{zfinaleq} and \eqref{lambdafinaleq}}. A detailed and comprehensive analysis of this study was conducted in \cite{Boiza:2024azh}. We next highlight the most important results for our current work.

To simulate the evolution of the universe described by our model, we numerically solve the system of equations after defining the initial conditions for the variables 
$x$, $y$, $z$ and $\lambda$. Our numerical integration begins at $\ln{(a_i/a_0)}=-15$, corresponding to a radiation-dominated epoch. At such an early stage, we assume that the contribution of dark energy is negligible and approximate the model using the $\Lambda$CDM framework. The initial value of $z$ is determined using $z_i=\Omega^{1/2}_{r\Lambda i}$, where $\Omega_{r\Lambda i}$ is given by

\begin{equation}\label{omegarcosmologicalconstant}
   \Omega_{r\Lambda i}=\frac{\Omega_{r0}(a_i/a_0)^{-4}}{\Omega_{r0}(a_i/a_0)^{-4}+\Omega_{m0}(a_i/a_0)^{-3}+\Omega_{\Lambda0}}.
\end{equation} 
Here, we adopt the parameters $\Omega_{m0}=0.3$, $\Omega_{r0}=8\times10^{-5}$ and $\Omega_{\Lambda0}=1-\Omega_{m0}-\Omega_{r0}$. We examine different initial values of $\phi_i$ within the range $5\times10^{-5}M_P$ to $5\times10^{-5/2}M_P$ to study the system's convergence to the tracking path. This range is selected to ensure two critical conditions: (i) the field starts far from the minimum $\phi/\eta=\pi$, and (ii) the system avoids a frozen state too close to the minimum. The variable $x$, which represents the kinetic energy contribution of the field, relates to $\dot{\phi}$ via a dimensionless form \eqref{xyvariables}. Initial values of $x_i$ are tested within the range $10^{-7}$ to $10^{-2}$, corresponding to a wide range of initial values for $\Omega_{\phi i}$. Finally, the variable $y$, which is linked to $\phi$ through the potential $V(\phi)$, also depends on the Hubble parameter $H$ \eqref{xyvariables}. At early-time, when the universe is radiation-dominated, $H$ can be approximated as

\begin{equation}\label{happrox}
   H_i\approx H_0\sqrt{\Omega_{r0}(a_i/a_0)^{-4}},
\end{equation}
where $H_0$ is the Hubble parameter today. Using these initial conditions, the evolution of the universe can be accurately simulated, capturing the interplay between radiation, matter, and the scalar field during different cosmic epochs. We refer the reader to \cite{Boiza:2024azh} for a detailed discussion of the initial conditions.

\begin{figure}
\centering
\subfigure[\label{fig:sfeosevolution}Evolution of $w_{\phi}$.]{\includegraphics[width = 0.485 \textwidth]{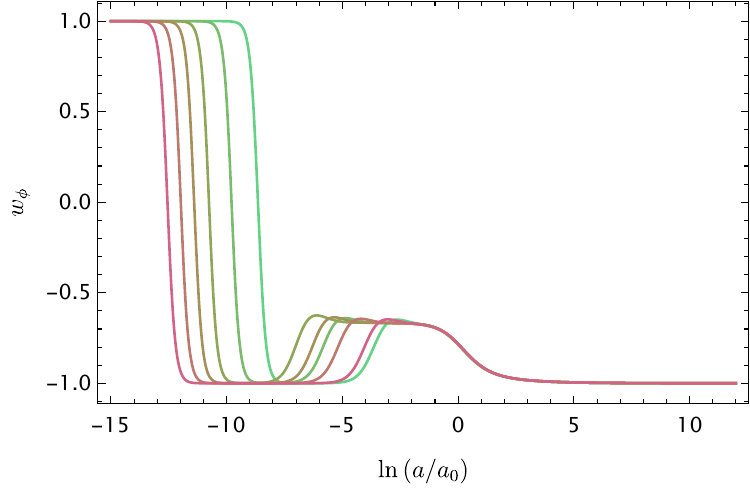}}~
\subfigure[\label{fig:omegaphievolution}Evolution of $\Omega_{\phi}$.]{\includegraphics[width = 0.5 \textwidth]{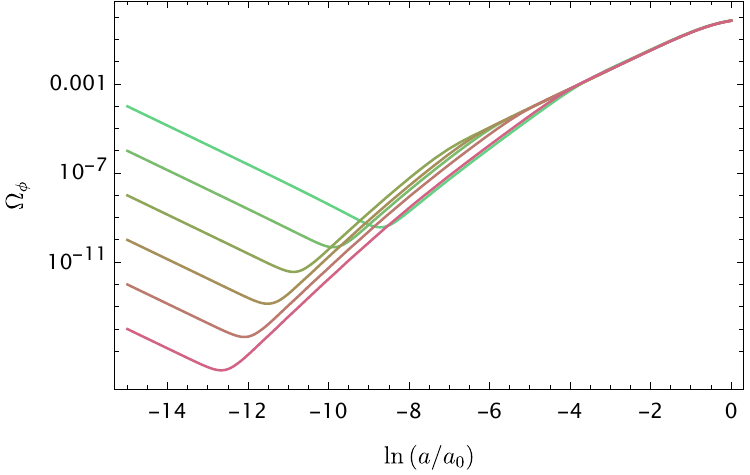}}
\caption{The left panel displays the evolution of the scalar field equation of state, $w_{\phi}$, for $n=0.5$, $\eta=M_P$. Each colour represents a distinct set of initial conditions used in computing the evolution, running from $\{\phi_i=5\times10^{-5/2}M_P, \: x_i=10^{-7}\}$ (lightest red) to $\{\phi_i=5\times10^{-5}M_P, \: x_i=10^{-2}\}$ (lightest green). The plot reveals an initial phase dominated by kinetic energy, where $w_{\phi}\approx1$. This phase is followed by a rapid transition into a frozen regime, characterised by $w_{\phi}\approx-1$, where the field temporarily ceases to evolve dynamically. After this period, the field enters the tracking phase, where it stabilises at $w_{\phi}\approx-2/3$, in agreement with the expectation from \eqref{sfeosgammamodel}. Ultimately, the field asymptotically approaches the de Sitter attractor, associated with $\lambda=0$, leading to $w_{\phi}=-1$. The right hand side panel shows the evolution of the scalar field density parameter $\Omega_{\phi}$ for $n=0.5$, $\eta=M_P$. Each colour corresponds to the same set of initial conditions as in the left hand side panel. The sets of initial conditions considered encompass a broad range of initial values for the scalar field density parameter, $\Omega_{\phi i}$, spanning from $10^{-14}$ to $10^{-4}$. The results indicate that, regardless of the initial conditions chosen for the calculation, the system consistently converges to a common tracking path. This behaviour demonstrates the robustness of the model, as the evolution of the density parameter $\Omega_{\phi}$ follows the same trajectory over time.}
\end{figure}

In Fig. \ref{fig:sfeosevolution}, we present the evolution of the scalar field equation of state, $w_{\phi}$, for the case $n=0.5$, $\eta=M_P$. The results confirm the tracking behaviour: following the frozen phase, where $w_{\phi}\approx-1$, all trajectories converge onto a common tracking path, regardless of their initial conditions. This universality in the evolution of $w_{\phi}$ demonstrates the attractor \COMRev{nature}\Rev{behaviour} of the tracking solution. The equation of state in the tracking regime can be analytically determined using \eqref{sfeosgammamodel}. For the specific case examined, $n=0.5$, we find $w_{\phi}\approx-2/3$. These theoretical predictions align well with the numerical results shown in the figure. A detailed analysis of the tracking behaviour in this model was carried out in \cite{Boiza:2024azh}.

In Fig. \ref{fig:omegaphievolution}, we illustrate how the selected range of initial conditions for $\phi$ and $\dot{\phi}$ spans a wide variety of values for the scalar field density parameter $\Omega_{\phi}$. At early-time, when the universe is dominated by radiation ($\Omega_{ri}\sim1$), we have also included scenarios where the subdominant components—matter and the scalar field—are comparable ($\Omega_{\phi i}\sim\Omega_{mi}$). This is particularly evident for the highest initial value of the scalar field density parameter, $\Omega_{\phi i}=10^{-4}$, as represented in the figure. From this maximum value, we progressively lower the scalar field densities toward values closer to those associated with a cosmological constant. This systematic reduction in $\Omega_{\phi i}$ demonstrates how the scalar field initial dominance can vary across orders of magnitude, yet the system consistently converges to the tracking path. Importantly, the convergence to the tracking path is shown to be independent of the initial conditions. Regardless of whether $\Omega_{\phi i}$ is relatively large or small at the outset, the system evolves predictably toward a common trajectory that aligns with the tracking solution. This robustness highlights the model's ability to address the coincidence problem, as it does not require fine-tuned initial conditions to achieve the observed late-time behaviour of the universe.

\begin{figure}
\centering
\includegraphics[width = 0.6\textwidth]{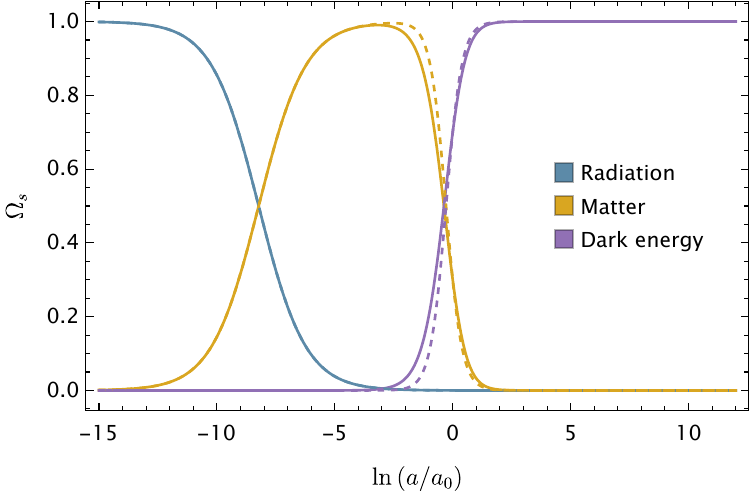}
\caption{\label{fig:omegasevolution}This panel illustrates the evolution of the density parameters $\Omega_s$ for different components of the universe: radiation ($s=r$, shown in blue), matter ($s=m$, shown in yellow), and dark energy ($s=\phi$, shown in purple). The solid lines depict the predictions of the proposed model for the case $n=0.5$, $\eta=M_P$, while the dashed lines correspond to a scenario where dark energy behaves as a cosmological constant. Notably, in this case, the evolution of the matter and dark energy components within our model deviates from the behaviour predicted in a cosmological constant scenario.}
\end{figure}

In Fig. \ref{fig:omegasevolution}, we illustrate the evolution of the density parameters, comparing them to a scenario where dark energy is modelled as a cosmological constant. We have set $n=0.5$ and $\eta=M_P$. The parameter $V_{\rm min}$ has been fixed in order to have $\Omega_{m}(a=a_0)\approx0.3$ and $\Omega_{r}(a=a_0)\approx8\times10^{-5}$ today. For our model, the matter density parameter $\Omega_m$ is slightly smaller during the matter-dominated era at late-time compared to the $\Lambda$CDM one. This reduction in $\Omega_m$ is accompanied by a corresponding increase in the scalar field density parameter $\Omega_{\phi}$. The reason for this behaviour can be traced to the equation of state of the scalar field during the tracking regime, as described by \eqref{sfeosgammamodel}. The scalar field equation of state $w_{\phi}$ deviates significantly from the cosmological constant value ($w_v=-1$) in the tracking regime. As a result, the scalar field plays a more prominent role during the matter-dominated era for larger $n$ values, leading to a slower decline in $\Omega_m$. For the specific case $n=0.5$, we find $w_{\phi}\approx-2/3$. Again, for a detailed discussion of this effect at background and perturbation levels, we refer the reader to \cite{Boiza:2024azh,Boiza:2024fmr}.

\section{Observational fit}
\label{sec:fit}

We utilise Cobaya's Monte-Carlo-Markov-Chain (MCMC) sampler \cite{Torrado:2020dgo, Lewis:2002ah, Lewis:2013hha, 2005math......2099N} and CAMB Boltzmann solver \cite{Lewis:1999bs,Howlett:2012mh} to generate the posterior distribution of the full cosmological parameters, including parameters for early-time CMB physics, i.e., $\Omega_{b0} h^2$, $\Omega_{c0} h^2$, $\theta_{\rm MC}$, $\ln A_s$, $n_s$ and $\tau_{\rm reio}$\footnote{The parameter $H_0 = 100 h$ is the Hubble constant derived from the characteristic acoustic peak angular diameter of CMB $\theta_{\rm MC}$ assuming a cosmic evolution where $H_0$ can be factored out as a scale \cite{Hu:1995en}. The model we consider adheres to this assumption. $\Omega_{b0}$ and $\Omega_{c0}$ are the density parameter of the baryon and the cold dark matter at the present time. We assume the minimal model of 3 neutrino generations with 1 massive neutrino of mass $0.06$eV and an effective number of $3.044$. The dark energy density $\Omega_{\Lambda0}$ can be derived following these assumptions. Following standard practice, the primordial perturbation spectrum is \COMRev{blue}\Rev{red}-tilted with the spectrum power $A_s$ and the spectrum tilt $1-n_s$. The reionization follows the standard $\tanh$ form with the optical depth $\tau_{\rm reio}$.}, for datasets that will be introduced in section~\ref{sec:data}. For each dataset, $3 \sim 5$ chains are deployed based on the early convergence test. Slow, non-convergent chains are dropped, leaving $2 \sim 4$ chains for statistical analysis. We utilise the stopping criteria of intra-chain and inter-chain Gelman-Rubin\footnote{The parameter R is defined as $R = \underset{\theta}{\max} \sqrt{1 + \left( \left< \left< \theta \right>_{\rm inner}^2 \right> - \left< \left< \theta \right>_{\rm inner} \right>^2 \right) \Big{/} \left< \left<\theta^2\right>_{\rm inner} - \left< \theta \right>_{\rm inner}^2 \right>}$, where $\theta$ are the parameters, $\left<\dots\right>_{\rm inner}$ is the average over an entire chain for inter-chain $R$, or over a segment of a chain for intra-chain $R$ with each segment $40N_\theta$ where $N_\theta$ is the number of parameters, and $\left<\dots\right>$ without subscript is the average over chains for inter-chain $R$ or over segments within a single chain for intra-chain $R$.} $R-1 < 0.1$.

To compare between different datasets and models, we primarily utilise 6 probes. The traditional Bayesian evidence $B$, deviance information criteria (DIC), and widely applicable information criteria (WAIC) \cite{10.5555/2567709.2502609} are for model comparison, while the Bayesian ratio $R$, the goodness of fit (GoF), and suspiciousness $S$ \cite{DES:2020hen, Raveri:2019gdp, PhysRevD.100.023512} are for data tension detection. For detailed explanation of each probe please refer to appendix~\ref{sec:stat}. To interpret the probes, all we need to know is that, except for the goodness of fit and suspiciousness, they follow Jefferys' scale, with larger value meaning statistically disfavoured and vice versa. For the goodness of fit and suspiciousness, we convert them into traditional $\sigma$-values.

Regarding the prior, we follow the standard practice of \cite{Planck2018} for the early-time cosmological parameters, as shown in table~\ref{table:prior}. Additional BBN (big bang nucleosynthesis) prior on the baryon density $\Omega_{b0} h^2$ and a cut $40 < H_0 < 100$ on the Hubble constant are imposed across all models and datasets considered in this work to stabilise chains without CMB data. For late-time physics, i.e., the dark energy model, we put the axion-like dark energy model in section~\ref{sec:model} with $\dot\phi (a_i / a_0 = 10^{-6}) = 0$ \Rev{(abbreviated as $\phi$CDM in figures) }against the $\Lambda$CDM model. The constraint on the initial field momentum is reasonable given the freezing property during radiation dominated era demonstrated in section~\ref{sec:model}.

Furthermore, a spatially flat FLRW universe is assumed, leading to a 6-parameter model for $\Lambda$CDM. For the axion-like dark energy model, the traditional approach of setting $V_{\rm min} = \rho_{\rm DE,0} \equiv 3 H^2 k^{-2} \Omega_{\Lambda0}$ is unsuitable since we may still be in the tracking regime at the current time, with a non-negligible amount of dark energy density excess over the potential minimum. As CAMB utilises forward-time integration for the background dark energy equation of motion, we impose a Gaussian likelihood of width $0.001/\sqrt{2}$ on $\mathcal{E}_{H_0} \equiv 1 - H_{0,{\rm EOM}}^2/ H_0^2$ reminiscent of an ad hoc Gaussian constraint, where $H_{0,{\rm EOM}}$ is the Hubble constant derived from the equation of motion and $H_0$ is the Hubble constant derived from $\theta_{\rm MC}$. By testing the constraint against the null hypothesis\footnote{Null hypothesis here refers to a model without the Gaussian constraint.}, the resulting non-normalised log Bayesian evidence of $1.03 \pm 0.11$, goodness of fit of $1.01 \pm 0.03\sigma$, and suspiciousness of $1.41 \pm 0.02\sigma$ (2 chains) confirm the lack of tension across the prior space\footnote{However, as we will notice later in Fig.~\ref{fig:fit_scatter_CMB} the finite width of the Gaussian does lead to under-representation of regions in the parameter space that are guaranteed to have $\mathcal{E}_{H_0} \sim 0$ given certain datasets. It would not affect our conclusion though, as the effect is not severe.}. We will subtract the likelihood of the constraint from all the likelihoods presented in this work, thus converting it into a prior probability\footnote{The associated prior volume for the Bayesian evidence in Eq.~\eqref{eq:BE} can be derived from the Bayesian ratio in the null hypothesis test.} of $\mathcal{E}_{H_0}$. With additional scalar action parameters of $n$, $\log_{10} ( V_{\rm min} / \rho_{\rm DE,0} )$, $\log_{10} (\eta / M_P)$ and the initial condition of $\log_{10} (\phi_i / \eta) \equiv \log_{10} \left( \phi (a_i / a_0 = 10^{-6}) / \eta \right)$, we may consider the axion-like dark energy as a (6+4)-parameter model, with one strongly constraint direction along $\mathcal{E}_{H_0}$.

\begin{table}[]
\centering
\begin{tabular}{|c|c|c c c|}
\hline
\multirow{6}{*}{$\Lambda$CDM parameters}
& $\Omega_{b0} h^2$                            & norm  & $0.0222$  & $0.0005$  \\
& $\Omega_{c0} h^2$                            & flat  & $0.001$   & $0.99$    \\
& $100\theta_{\rm MC}$                     & flat  & $0.5$     & $10$      \\
& $\ln ( 10^{10} A_s )$                     & flat  & $1.61$    & $3.91$    \\
& $n_s$                                     & flat  & $0.8$     & $1.2$     \\
& $\tau_{\rm reio}$                        & flat  & $0.01$    & $0.8$     \\
\hline
\multirow{4}{*}{Extended parameters}
& $n$                                       & flat  & $0.0001$  & $1$       \\
& $\log_{10} (V_{\rm min}/\rho_{\rm DE,0})$ & flat  & $-2$      & $0$       \\
& $\log_{10} \eta / M_P$                    & flat  & $-2$      & $1$       \\
& $\log_{10} (\phi_i / \eta)$               & flat  & $-4.5$    & $0.5$     \\
\hline
\multirow{2}{*}{Constraint}
& $\mathcal{E}_{H_0}$                & norm  & $0$       & $0.001/\sqrt{2}$  \\
& $H_0$                & flat  & $40$       & $100$  \\
\hline
\end{tabular}
\caption{\label{table:prior}
The model parameters of $\Lambda$CDM model and the axion-like dark energy model considered in this work. For the normal distribution two columns provide the mean and the width respectively, and for the flat distribution two columns corresponds to the minimum and the maximum. }
\end{table}

While there are prior works \cite{Hossain:2023lxs, Hossain:2025grx} on the cosmological fit for the axion-like dark energy model considered in this work, the dataset and the prior selected are substantially different. For naturalness and our interest in the tracking behaviour of the model, the model parameter $\eta$ is chosen to be around $O(M_P)$. This choice of $\eta$ was rarely explored as it deviates further from the standard-model-inspired axion model
. However, as shown previously \cite{Boiza:2024azh}, coincidentally the tracking regime ends at the current era only when $\eta = O(M_P)$, and only during which the scalar field deviates from $w = -1$, altering the cosmic evolution.

On the other hand, priors on $n$ and $\phi_i$ are chosen for non-physical reasons. One major reason we only consider $n \leq 1$ (no overlap with \cite{Hossain:2023lxs}) is that a greater $n$ steepens the potential, to the point of inducing numerical instability. Although this should be resolvable by the field redefinition of Eq.~\eqref{lambdaphimodel}, or at least eased slightly with a better defined prior space such as $\log_{10} \Omega_\phi (a_i/ a_0 = 10^{-6})$ instead of $\log_{10} (\phi_i / \eta)$, we choose to stick with the action parameters, as it is clearly shown in Eq.~\eqref{sfeosgammamodel} that a greater $n$ only leads to higher $w$ in the tracking region and thus a stronger tension with current observations. Meanwhile, the prior chosen on $V_{\rm min}$ definitely suffices with the prior on $n$.

This section is divided into 3 subsections. Section \ref{sec:data} explains the dataset considered in this paper. Section \ref{sec:result} presents the main result of this work, i.e, the cosmological parameters of the axion-like dark energy model under various datasets. Section \ref{sec:evolution} discusses the effect of the axion-like dark energy on the cosmic evolution.

\subsection{Data}
\label{sec:data}

We consider multiple astrophysical and cosmological datasets. For simplicity, they are grouped together according to the theoretical origin as CMB (Planck2018 low-$\ell$ TTEE \cite{Planck:2019nip}, NPIPE CamSpec high-$\ell$ TTTEEE \cite{Rosenberg:2022sdy}, PR4 CMB lensing \cite{Carron:2022eyg, Carron:2022eum}) , BAO (DESI DR1 without full-shape \cite{DESI:2024mwx, DESI:2024lzq, DESI:2024uvr}), SNe (Pantheon+ SNe catalog \cite{Brout:2022vxf} without anchoring of SNe standardised absolute magnitude), low-$z$ (Riess et al. low-$z$ anchors \cite{Riess:2020fzl}), and DES Y1 morphological datasets \cite{DES:2017myr}. Additional $f\sigma_8$ dataset (table 2 of \cite{Avila:2022xad}, with data entries from \cite{eBOSS:2020lta, Turnbull:2011ty, Achitouv:2016mbn, Beutler:2012px, Feix:2015dla, BOSS:2016wmc, BOSS:2013eso, Blake:2012pj, Nadathur:2019mct, BOSS:2013mwe, Wilson:2016ggz, eBOSS:2018yfg, Okumura:2015lvp}), SDSS DR7 LRG dataset \cite{Reid:2009xm} and eBOSS DR14 Ly-$\alpha$ dataset \cite{eBOSS:2018qyj} are presented in \COMRev{our figures in section~\ref{sec:evolution}, but are excluded from our cosmological fit}\Rev{Figs.~\ref{fig:fsigma8_CMB}, \ref{fig:fsigma8_CMB_BAO} and \ref{fig:matter_power_spectrum} in section~\ref{sec:evolution} for visual comparison. They are not considered in the MCMC analysis to avoid potential overlap with the DESI BAO dataset}.

We do not consider the distance priors of CMB \cite{Chen:2018dbv} as in \cite{Hossain:2025grx}, and instead rely on CAMB to capture as much information as possible. This is crucial for our model as it exhibits non-trivial tracking behaviour, and with small enough initial field strength it is possible to have non-negligible initial dark energy density, as shown in Fig.~\ref{fig:omegaphievolution}. The difference in the CMB dataset will manifest itself throughout the entire analysis, leading to different parameter estimates, model statistics, and eventually the conclusion drawn. With Planck as our base dataset, we then progressively add in later-time datasets in the following order: BAO, SNe, low-$z$, and DES Y1. For the evaluation of the Bayesian ratio and suspiciousness between each dataset, we also consider 3 smaller datasets: BAO, SNe+low-$z$ and DES Y1.

Due to time constraint, even though the Bayesian evidence, DIC, WAIC, etc. have stabilised, we cannot meet the stopping criteria when fitting DES Y1 dataset with the axion-like dark energy model (inter-chain $R-1 \sim 0.8$ for $\log_{10} (V_{\rm min}/\rho_{\rm DE,0})$ as the model exhibits bimodal behaviour, with 3 chains stuck at one side and only 2 successfully migrating between two branches), and one should be cautious against reading too much into its posterior distribution.

\subsection{Cosmological parameters}
\label{sec:result}

For the visualisation of the cosmological fit, we utilise GetDist package \cite{Lewis:2019xzd} and choose a different set of parameters than the parameter set that leads to a flat prior space in table~\ref{table:prior}. For the early-time CMB physics we stick with the well-accepted 6-dimensional parameter space of $\Omega_{b0} h^2$, $\Omega_{c0} h^2$, $100\theta_{\rm MC}$, $\ln (10^{10} A_s)$, $n_s$ and $\tau_{\rm reio}$. For the late-time physics, we choose the standard observables of Hubble constant $H_0$, total matter density parameter at current time $\Omega_{m0}$, the matter spectrum power $\sigma_8$ at the scale of $8{\rm Mpc}/h$, and the rescaled matter spectrum power $S_8 = \sigma_8\sqrt{\Omega_{m0} / 0.3}$.

For the model parameters of the axion-like dark energy, \COMRev{we choose the excess dark energy density over the potential minimum $\log_{10} (\rho_{\rm DE,0} / V_{\rm min} - 1)$ and the curvature of the potential around the minimum $\log_{10} (m^2 / V_{\rm min}) = \log_{10} (n \eta^{-2} / 2)$ defined in Eq.~\eqref{minimumexpansion} instead of $\log_{10} (V_{\rm min} / \rho_{\rm DE,0})$ and $\log_{10}(\eta / M_P)$, as these two are directly tied to the late-time observables. The prior of $\log_{10} (\rho_{\rm DE,0} / V_{\rm min} - 1)$ would be non-flat though, with a density that penalises negative values linearly in terms of log prior probability}\Rev{instead of the field-theory-inspired flat prior space of $\log_{10} ( V_{\rm min} / \rho_{\rm DE,0} )$ and $\log_{10} ( \eta / M_P )$, we re-parametrise them as the excess dark energy density over the potential minimum $\log_{10} ( \rho_{\rm DE,0} / V_{\rm min} - 1)$ and the curvature of the potential around the minimum $\log_{10} ( m^2 / V_{\rm min} ) = \log_{10} ( n \eta^{-2} / 2 )$ defined in Eq.~\eqref{minimumexpansion}, as these two are directly tied to the late-time observables. Effectively, we arrive at a prior probability density in $\log_{10} ( \rho_{\rm DE,0} / V_{\rm min} - 1)$ space that is exponentially suppressed toward the negative direction}.

With our choice of the parameters fixed, let us first analyse if the axion-like dark energy model could affect the early-time CMB physics, thus leading to different late-time predictions that may ease some cosmological tension. To achieve this goal, we consider a series of MCMC analyses that gradually includes CMB, BAO, SNe, low-$z$ and DES Y1 datasets.

The mean and the standard deviation of the cosmological parameters, late-time observables, and statistical probes of these analyses are shown in table~\ref{table:result_LambdaCDM} for $\Lambda$CDM model and in table~\ref{table:result_scalar} for the axion-like dark energy model. From the tables, one can infer that the model we consider in this work does not provide a statistically significant benefit, as most statistical probes very weakly but consistently prefer $\Lambda$CDM model over axion-like dark energy model\footnote{We draw different conclusion than \cite{Hossain:2025grx} as different CMB datasets are utilised. Notice that IC values in our work differ by a factor of 2 compared to \cite{Hossain:2025grx}.}. The statistical preference is supported by the similarity between the predicted values of $H_0$ and $S_8$ by $\Lambda$CDM model and the axion-like dark energy model, suggesting that neither Hubble tension nor $S_8$ tension are solved.

\begin{table}
\centering
\scriptsize
\begin{tabular}{|c||c|c|c|c|c|}
\hline
                            & CMB                   & + BAO                 & + SNe                 & + low-$z$             & + DES Y1  \\
\hhline{|=#=|=|=|=|=|}
$10^3 \Omega_{b0} h^2$ & $22.19    \pm0.13$    & $22.29    \pm0.12$    & $22.26    \pm0.12$    & $22.34    \pm0.12$    & $22.38    \pm0.12$    \\
$10^3 \Omega_{c0} h^2$ & $119.7    \pm1.0$     & $118.26   \pm0.81$    & $118.56   \pm0.79$    & $117.80   \pm0.76$    & $117.27   \pm0.72$    \\
$10^5\theta_{\rm MC}$   & $1040.77  \pm0.25$& $1040.94  \pm0.24$    & $1040.91  \pm0.24$    & $1041.01  \pm0.24$    & $1041.05  \pm0.24$    \\
$\ln (10^{10} A_s)$ & $3.037    \pm0.014$   & $3.044    \pm0.014$   & $3.043    \pm0.014$   & $3.047    \pm0.014$   & $3.046    \pm 0.014$  \\
$n_s$               & $0.9636   \pm0.0040$  & $0.9672   \pm0.0036$  & $0.9665   \pm0.0036$  & $0.9685   \pm0.0036$  & $0.9693   \pm0.0036$  \\
$\tau_{\rm reio}$   & $0.0524   \pm0.0071$  & $0.0571   \pm0.0071$  & $0.0562   \pm0.0070$  & $0.0587   \pm0.0071$  & $0.0589   \pm0.0071$  \\
\hline
$H_0$               & $67.24    \pm0.46$    & $67.89    \pm0.36$    & $67.76    \pm0.35$    & $68.12    \pm0.34$    & $68.36    \pm0.32$    \\
$\Omega_{m0}$          & $0.3154   \pm0.0064$  & $0.3064   \pm0.0048$  & $0.3082   \pm0.0047$  & $0.3034   \pm0.0044$  & $0.3003   \pm0.0042$  \\
$\sigma_8$          & $0.8077   \pm0.0055$  & $0.8064   \pm0.0056$  & $0.8067   \pm0.0056$  & $0.8061   \pm0.0057$  & $0.8041   \pm0.0055$  \\
$S_8$               & $0.828    \pm0.011$   & $0.8149   \pm0.0090$  & $0.8176   \pm0.0090$  & $0.8107   \pm0.0087$  & $0.8044   \pm0.0080$  \\
\hline
DIC                 & $5497.90  \pm0.12$    & $5507.21  \pm0.37$    & $6209.34  \pm0.14$    & $6217.77  \pm0.43$    & $6477.38  \pm0.28$    \\
WAIC                & $5499.09  \pm0.50$    & $5507.85  \pm0.18$    & $6210.36  \pm0.21$    & $6218.68  \pm0.21$    & $6481.16  \pm0.22$    \\
$-\ln B$            & $5499.3   \pm1.1$     & $5508.13  \pm0.73$    & $6210.51  \pm0.44$    & $6219.1   \pm1.4$     & $6479.9   \pm1.2 $    \\
\hline
\end{tabular}
\caption{\label{table:result_LambdaCDM} Mean and standard deviation of cosmological parameters, late-time observables, and statistical probes for $\Lambda$CDM model. From left to right are gradually larger datasets that progressively add in datasets of CMB, BAO, etc., as defined in section~\ref{sec:data}.}
\end{table}

\begin{table}
\centering
\scriptsize
\begin{tabular}{|c||c|c|c|c|c|}
\hline
                            & CMB                   & + BAO                 & + SNe                 & + low-$z$             & + DES Y1  \\
\hhline{|=#=|=|=|=|=|}
$10^3 \Omega_{b0} h^2$ & $22.18    \pm0.13$    & $22.28    \pm0.12$    & $22.28    \pm0.12$    & $22.34    \pm0.12$    & $22.39    \pm0.12$    \\
$10^3 \Omega_{c0} h^2$ & $119.8    \pm1.1$     & $118.29   \pm0.80$    & $118.40   \pm0.83$    & $117.73   \pm0.79$    & $117.22   \pm0.73$    \\
$10^5\theta_{\rm MC}$   & $1040.76  \pm0.25$& $1040.95  \pm0.24$    & $1040.93  \pm0.24$    & $1041.03  \pm0.23$    & $1041.06  \pm0.26$    \\
$\ln (10^{10} A_s)$ & $3.039    \pm0.014$   & $3.044    \pm0.014$   & $3.044    \pm0.014$   & $3.048    \pm0.014$   & $3.048    \pm0.014$   \\
$n_s$               & $0.9634   \pm0.0040$  & $0.9671   \pm0.0036$  & $0.9669   \pm0.0037$  & $0.9686   \pm0.0036$  & $0.9694   \pm0.0035$  \\
$\tau_{\rm reio}$   & $0.0532   \pm0.0071$  & $0.0568   \pm0.0073$  & $0.0566   \pm0.0071$  & $0.0590   \pm0.0072$  & $0.0596   \pm0.0073$  \\
\hline
$\log_{10}(\rho_{\rm DE,0}/V_{\rm min}-1)$  & $\color{BrickRed}{\color{black}-2.9}^{+2.1}_{\color{black}-0.7}$  & $\color{Cyan}{\color{black}-2.92}^{+0.45}_{\color{black}-0.75}$   & $\color{BrickRed}{\color{black}-3.0}^{+2.6}_{\color{black}-0.6}$  & $\color{Cyan}{\color{black}-2.94}^{+0.43}_{\color{black}-0.76}$   & $\color{BrickRed}{\color{black}-3.0}^{+3.1}_{\color{black}-0.8}$  \\
$\log_{10}(\eta/M_P)$                       & $\color{cyan}{\color{black}-1.25}^{+0.31}_{\color{white}-?}$      & $\color{cyan}{\color{black}-2.0}^{+0.8}_{\color{white}-?}$        & $\color{cyan}-0.8$    & $\color{cyan}{\color{black}-2.0}^{+0.7}_{\color{white}-?}$    & $\color{cyan}-0.8$    \\
$n$                                 & $0^{\color{Cyan}+0.6}$    & $0^{\color{Cyan}+0.6}$    & $0^{\color{Cyan}+0.5}$    & $0^{\color{Cyan}+0.6}$    & $0^{\color{Cyan}+0.6}$    \\
$\log_{10}\left(\phi_i/\eta\right)$ & $0.5_{\color{Cyan}-3.3}$  & $0.5_{\color{Cyan}-3.0}$  & $0.5_{\color{Cyan}-2.0}$  & $0.5_{\color{Cyan}-2.9}$  & $0.5_{\color{Cyan}-1.5}$  \\
\hline
$H_0$           & $\color{BrickRed}{\color{black}67.05}^{+0.62}_{\color{blue}-0.77}$    & $67.77    \pm0.39$    & $67.62    \pm0.45$    & $68.08    \pm0.36$    & $68.31    \pm0.33$    \\
$\Omega_{m0}$      & $\color{blue}{\color{black}0.3177}^{+0.0095}_{\color{BrickRed}-0.0078}$  & $0.3075   \pm0.0049$  & $0.3091   \pm0.0053$  & $0.3036   \pm0.0046$  & $0.3006   \pm0.0042$  \\
$\sigma_8$  & $\color{BrickRed}{\color{black}0.8056}^{+0.0070}_{\color{blue}-0.0079}$   & $0.8049   \pm0.0062$  & $0.8042   \pm0.0067$  & $0.8052   \pm0.0059$  & $0.8035   \pm0.0059$  \\
$S_8$               & $0.831    \pm0.012$   & $0.8149   \pm0.0091$  & $0.8163   \pm0.0090$  & $0.8100   \pm0.0089$  & $0.8042   \pm0.0080$  \\
\hline
DIC                 & $5498.13  \pm0.54$    & $5507.87  \pm0.13$    & $6210.08  \pm0.35$    & $6218.58  \pm0.01$    & $6478.21  \pm0.53$    \\
WAIC                & $5499.03  \pm0.53$    & $5508.42  \pm0.15$    & $6210.85  \pm0.03$    & $6219.12  \pm0.29$    & $6481.97  \pm0.92$    \\
$-\ln B$            & $5499.74  \pm0.99$    & $5509.2   \pm2.4$     & $6210.17  \pm0.04$    & $6219.09  \pm0.20$    & $6482.8   \pm2.8$     \\
\hline
$\Delta$DIC         & $   0.23  \pm0.59$    & $   0.66  \pm0.41$    & $   0.74  \pm0.39$    & $   0.81  \pm0.51$    & $   0.83  \pm0.63$    \\
$\Delta$WAIC        & $  -0.06  \pm0.77$    & $   0.56  \pm0.24$    & $   0.48  \pm0.45$    & $   0.43  \pm0.39$    & $   0.81  \pm0.97$    \\
$-\Delta\ln B$      & $   0.4   \pm1.6$     & $   1.0   \pm2.4$     & $  -0.35  \pm0.52$    & $  -0.1   \pm1.7$     & $   2.9   \pm3.2$     \\
\hline
\end{tabular}
\caption{\label{table:result_scalar} Mean and standard deviation of cosmological parameters, late-time observables, and statistical probes for the axion-like dark energy model in section~\ref{sec:model}. From left to right are gradually larger datasets of CMB, CMB + BAO, CMB + BAO + SNe, etc., as defined in section~\ref{sec:data}. $\Delta{\rm ICs}$ are with respect to $\Lambda$CDM model presented in table~\ref{table:result_LambdaCDM}. For parameters not following Gaussian distribution we provide the median and 68\% lower and upper bounds (if valid) instead, with color coding for how heavy the tail is ({\color{BrickRed}red} for short tail, black for Gaussian, {\color{blue}blue} for exponential, and {\color{Cyan}cyan} for long tail.) For single-sided distributions we report the modal and the single-sided 68\% bound instead.}
\end{table}

\begin{figure}
\centering
\includegraphics[width = 0.96\textwidth]{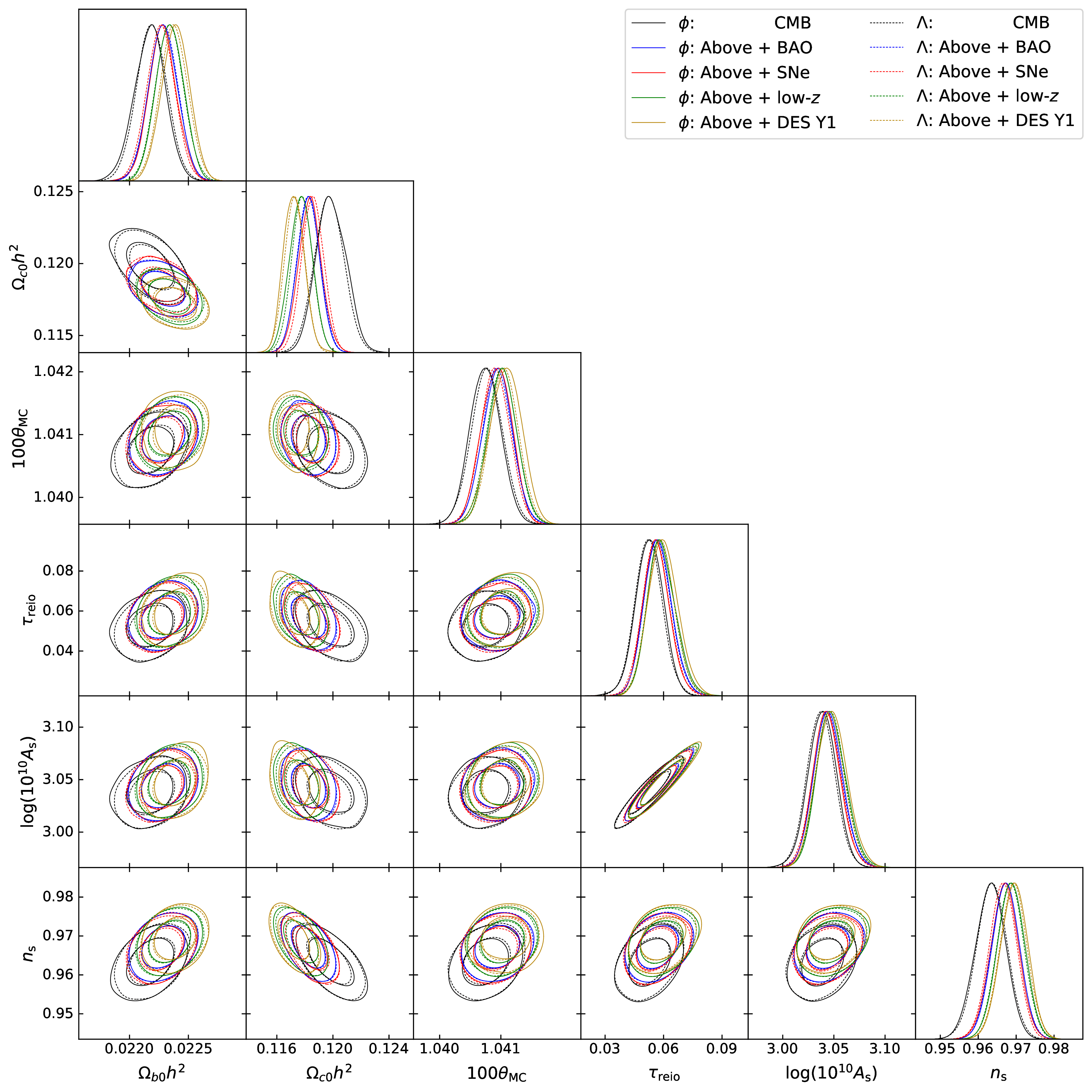}
\caption{\label{fig:fit_CMBplus_early_params} 68\% and 95\% C.L. posterior distribution of the early-time parameters from CMB (Black), CMB + BAO ({\color{RoyalBlue}blue}), CMB + BAO + SNe ({\color{BrickRed}red}), CMB + BAO + SNe + low-$z$ ({\color{ForestGreen} green}), and CMB + BAO + SNe + low-$z$ + DES Y1 ({\color{Brown}Brown}) data, with $\Lambda$CDM model the \COMRev{lighter shade}\Rev{dashed} contours and the axion-like dark energy model \COMRev{in section~\ref{sec:model}}\Rev{($\phi$CDM)} the \COMRev{darker shade}\Rev{solid} contours. The result demonstrates the stability of the early-time physics\COMRev{ while strongly indicates the inability of the axion-like dark energy model to ease the tension between astrophysical data of low-$z$ and cosmological data of CMB + BAO}.}
\end{figure}

\begin{figure}
\centering
\includegraphics[width = 0.64\textwidth]{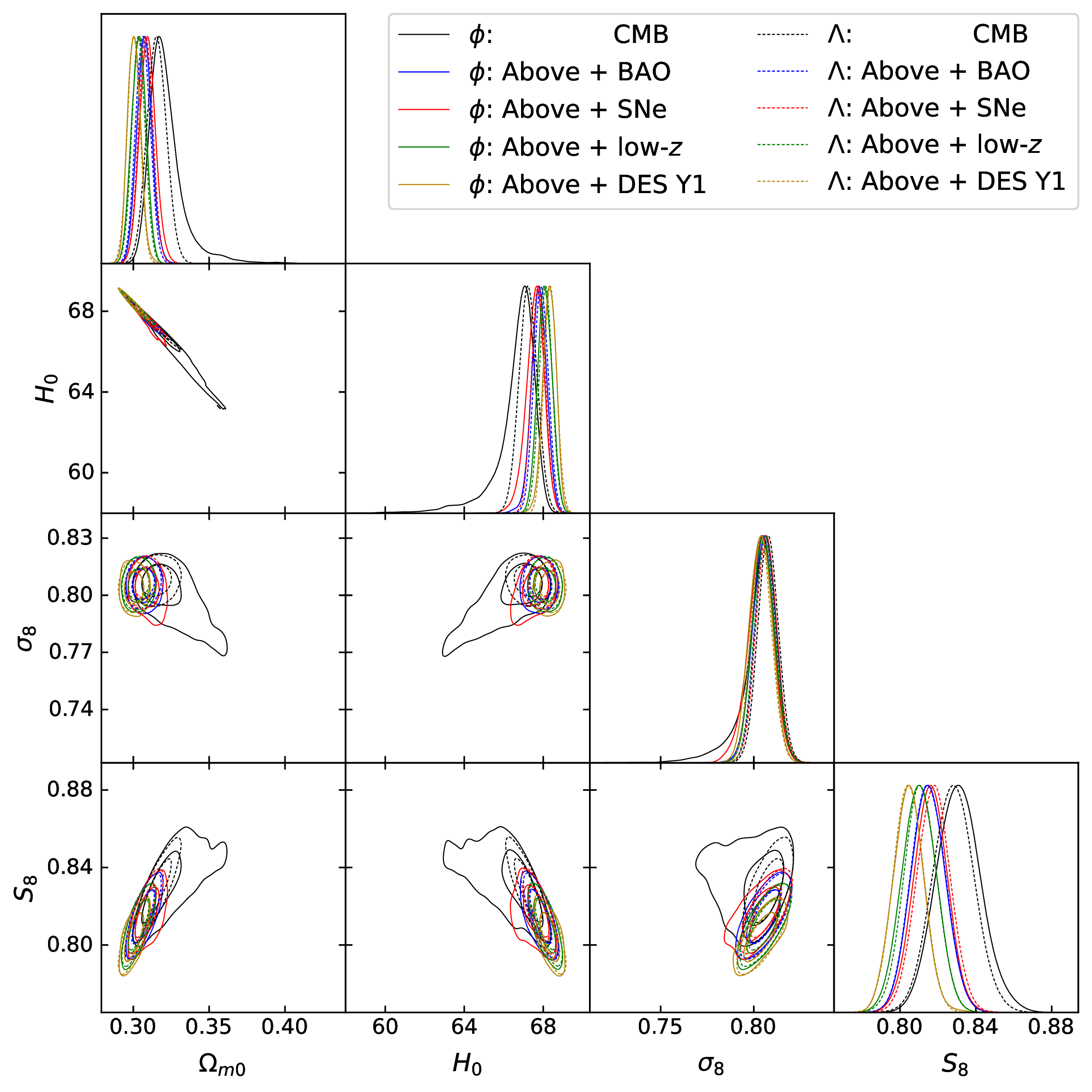}
\caption{\label{fig:fit_CMBplus_late_params} 68\% and 95\% C.L. posterior distribution of the late-time quantities from CMB (Black), CMB + BAO ({\color{RoyalBlue}blue}), CMB + BAO + SNe ({\color{BrickRed}red}), CMB + BAO + SNe + low-$z$ ({\color{ForestGreen} green}), and CMB + BAO + SNe + low-$z$ + DES Y1 ({\color{Brown}Brown}) data, with $\Lambda$CDM model the \COMRev{lighter shade}\Rev{dashed} contours and the axion-like dark energy model \COMRev{in section~\ref{sec:model}}\Rev{($\phi$CDM)} the \COMRev{darker shade}\Rev{solid} contours. The result confirms the inability of the axion-like dark energy model to increase the predicted $H_0$ value. \Rev{The non-Gaussian shape of the contours is a feature of the model, as explained in appendix~\ref{sec:prior}.}}
\end{figure}

Furthermore, as shown in Figs.~\ref{fig:fit_CMBplus_early_params} and \ref{fig:fit_CMBplus_late_params}, the posterior distribution of the cosmological parameters also remains identical, with the sole exception of an extension toward lower value of $H_0$ for the axion-like dark energy model in the case of CMB dataset. This does not help with the Hubble tension, and is expected from a dark energy model with $w > -1$ \cite{Roy:2022fif}.

However, \COMRev{Figs.~\ref{fig:fit_CMBplus_late_params} only show the inability of the axion-like dark energy model to modify CMB predictions, which is half of the story. To confirm the persistence of the tension in the axion-like dark energy model}\Rev{the main goal of the model is not to alter the CMB prediction significantly, but to provide a dynamical scalar alternative to the cosmological constant. In this regard}, we perform the MCMC analyses of BAO, SNe+low-$z$ and DES Y1 datasets alone without CMB dataset. Mean and standard deviation of the cosmological parameters, late-time observables, and statistical probes are shown in table~\ref{table:tension_LambdaCDM} for the $\Lambda$CDM model and in table~\ref{table:tension_scalar} for the axion-like dark energy model. Notice that the lack of constraint on $\sigma_8$ and $S_8$ for BAO and SNe + low-$z$ datasets is expected, as we impose no a priori constraint on the primordial perturbation. Similarly, $n_s$ and $\tau_{\rm reio}$ are omitted due to the lack of constraint by the late-time physics. $\Omega_{b0} h^2$ is omitted as well since it would only reflect the BBN prior. In addition, the DES Y1 posterior for the axion-like dark energy model is a rough estimate, as we cannot reach the stopping criteria.

\begin{table}
\centering
\scriptsize
\begin{tabular}{|c||c|c|c|}
\hline
                    & BAO                   & SNe + low-$z$         & DES Y1                \\
\hhline{|=#=|=|=|}
$10^3 \Omega_{c0} h^2$ & $116.7    \pm8.3$     & $155      \pm11$      & $\color{blue}{\color{black}86}^{+23}_{\color{black}-10}$  \\
$10^5\theta_{\rm MC}$   & $1042 \pm10$      & $1090     \pm11$      & $1005^{+45}_{-34}$    \\
$\ln (10^{10} A_s)$ &                       &                       & $3.41^{+0.30}_{-0.26}$ \\
\hline
$H_0$               & $68.71    \pm0.80$    & $73.2     \pm1.3$     & $68.2     \pm6.9$     \\
$\Omega_{m0}$          & $0.295    \pm0.015$   & $0.332    \pm0.018$   & $\color{blue}{\color{black}0.250}^{+0.040}_{\color{black}-0.032}$   \\
$\sigma_8$          &                       &                       & $0.858    \pm0.084$   \\
$S_8$               &                       &                       & $0.796    \pm0.028$   \\
\hline
DIC                 & $ 8.45    \pm0.05$    & $703.37   \pm0.03$    & $260.64   \pm0.75$    \\
WAIC                & $ 8.51    \pm0.15$    & $703.31   \pm0.03$    & $262.63   \pm0.10$    \\
$-\ln B$            & $ 8.86    \pm0.34$    & $703.49   \pm0.11$    & $262.35   \pm0.87$    \\
\hline
Tension against     & CMB                   & CMB + BAO             & CMB + BAO + SNe + low-$z$ \\
\hline
$-\ln R$            & $ 0.0     \pm1.4$     & $ 7.5     \pm1.6$     & $-1.6     \pm2.1$     \\
GoF                 & $ 1.94 \pm0.30\sigma$ & $ 4.30 \pm0.20\sigma$ & $ 2.54 \pm0.30\sigma$ \\
$S$                 & $ 1.68 \pm0.18\sigma$ & $ 4.06 \pm0.17\sigma$ & $ 1.60 \pm0.07\sigma$ \\
\hline
\end{tabular}
\caption{\label{table:tension_LambdaCDM} Mean and standard deviation of cosmological parameters, late-time observables, and statistical probes for $\Lambda$CDM model. From left to right are datasets of BAO, SNe + low-$z$, and DES Y1. Tension probes of $-\ln R$, GoF and $S$ are with respect to $\Lambda$CDM model inside table~\ref{table:result_LambdaCDM} according to ``Tension against'' row. For parameters not following Gaussian distribution we provide the median and 68\% lower and upper bounds (if valid) instead, with colour coding for how heavy the tail is ({\color{BrickRed}red} for short tail, black for Gaussian, {\color{blue}blue} for exponential, and {\color{Cyan}cyan} for long tail.)}
\end{table}

\begin{table}
\centering
\scriptsize
\begin{tabular}{|c||c|c|c|}
\hline
                            & BAO                   & SNe + low-$z$         & DES Y1                \\
\hhline{|=#=|=|=|}
$10^3 \Omega_{c0} h^2$     & $111  \pm11$  & $116      \pm29$              & $107^{+15}_{\color{blue}-19}$ \\
$10^5 \theta_{\rm MC}$  & $1036 \pm12$  & $1081^{+15}_{\color{blue}-18}$& $\color{red}{\color{black}1048}^{+36}_{\color{black}-38}$      \\
$\ln ( 10^{10} A_s )$   &               &                               & $\color{black}{\color{black}3.47}^{+0.27}_{\color{black}-0.29}$  \\
\hline
$\log(\rho_{\rm DE,0}/V_{\rm min}-1)$   & $\color{BrickRed}{\color{black}-2.8}^{+2.1}_{\color{black}-0.6}$  & $-0.08^{+0.80}_{\color{blue}-0.95}$                           & $\color{cyan}{\color{black}-1.0}^{+1.8}_{\color{blue}-1.6}$  \\
$\log(\eta/M_P)$                        & $\color{blue}{\color{black}-0.89}^{+0.56}_{\color{white}-0.85}$   & $\color{white}{\color{black}0.5}^{+0.2}_{\color{blue}-1.0}$   & $\color{white}{\color{black}-0.3}^{+?}_{\color{cyan}-1.0}$   \\
$n$                             & $\color{Cyan} 0.4$        & $0^{\color{Cyan}+0.5}$& $\color{Cyan} 0.4$        \\
$\log\left(\phi_i/\eta\right)$  & $0.5_{\color{Cyan}-3.2}$  & $\color{Cyan}-1.4$    & $0.5_{\color{Cyan}-2.9}$  \\
\hline
$H_0$   & $\color{BrickRed}{\color{black}68.5}^{+1.0}_{\color{Cyan}-1.4}$   & $73.0 \pm1.3$ & $70.4 \pm5.5$ \\
$\Omega_{m0}$      & $0.291\pm0.016$   & $0.281^{+0.043}_{\color{blue}-0.053}$ & $0.266    \pm0.046$   \\
$\sigma_8$                  &                       &                       & $0.862    \pm0.093$   \\
$S_8$                       &                       &                       & $0.802    \pm0.031$   \\
\hline
DIC                         & $ 8.53    \pm0.12$    & $703.54   \pm0.33$    & $260.86   \pm0.72$    \\
WAIC                        & $ 8.26    \pm0.21$    & $703.40   \pm0.45$    & $262.36   \pm0.68$    \\
$-\ln B$                    & $ 8.33    \pm0.26$    & $703.56   \pm0.11$    & $262.26   \pm0.29$    \\
\hline
$\Delta$DIC                 & $ 0.08    \pm0.14$    & $  0.17   \pm0.34$    & $  0.2    \pm1.1$     \\
$\Delta$WAIC                & $-0.24    \pm0.27$    & $  0.09   \pm0.46$    & $ -0.27   \pm0.69$    \\
$-\Delta\ln B$              & $-0.53    \pm0.48$    & $  0.07   \pm0.17$    & $ -0.10   \pm0.98$    \\
\hline
Tension against             & CMB                   & CMB + BAO             & CMB + BAO + SNe + low-$z$ \\
\hline
$R$                         & $  1.2    \pm2.8$     & $ 6.3     \pm2.8$     & $ 1.5     \pm2.8$     \\
GoF                         & $ 2.33 \pm0.25\sigma$ & $ 4.20 \pm0.23\sigma$ & $ 2.93 \pm0.35\sigma$ \\
$S$                         & $ 2.08 \pm0.41\sigma$ & $ 3.96 \pm0.26\sigma$ & $ 1.86 \pm0.47\sigma$ \\
\hline
$\Delta R$                  & $  1.2    \pm3.1$     & $-1.2     \pm3.2$     & $ 3.1     \pm3.5$     \\
$\Delta{\rm GoF}$           & $ 0.39 \pm0.39\sigma$ & $-0.10 \pm0.30\sigma$ & $ 0.39 \pm0.46\sigma$ \\
$\Delta S$                  & $ 0.40 \pm0.45\sigma$ & $-0.10 \pm0.31\sigma$ & $ 0.26 \pm0.48\sigma$ \\
\hline
\end{tabular}
\caption{\label{table:tension_scalar} Mean and standard deviation of cosmological parameters, late-time observables, and statistical probes for the axion-like dark energy model in section~\ref{sec:model}. From left to right are datasets of BAO, SNe + low-$z$, and DES Y1. $\Delta{\rm ICs}$ and delta of tension probes are with respect to $\Lambda$CDM model presented in table~\ref{table:tension_LambdaCDM}. Tension probes of $-\ln R$, GoF and $S$ are with respect to axion-like dark energy model inside table~\ref{table:result_scalar} according to ``Tension against'' row. For parameters not following Gaussian distribution we provide the median and 68\% lower and upper bounds (if valid) instead, with colour coding for how heavy the tail is ({\color{BrickRed}red} for short tail, black for Gaussian, {\color{blue}blue} for exponential, and {\color{Cyan}cyan} for long tail.) If the distribution is clearly single-sided we report the modal and the single-sided 68\% bound instead.}
\end{table}

Following the statistical analysis described at the beginning of section~\ref{sec:fit}, we carry out the statistical probes of Bayesian ratio, goodness of fit and suspiciousness, and find that the axion-like dark energy model overall shows no statistical (dis)-advantage over $\Lambda$CDM model in terms of data tension\Rev{, demonstrating the potential of the dynamical, axion-like dark energy model as an alternative to the cosmological constant}.

\begin{figure}
\centering
\includegraphics[trim = 0 0 0 450, clip, width = 0.64\textwidth]{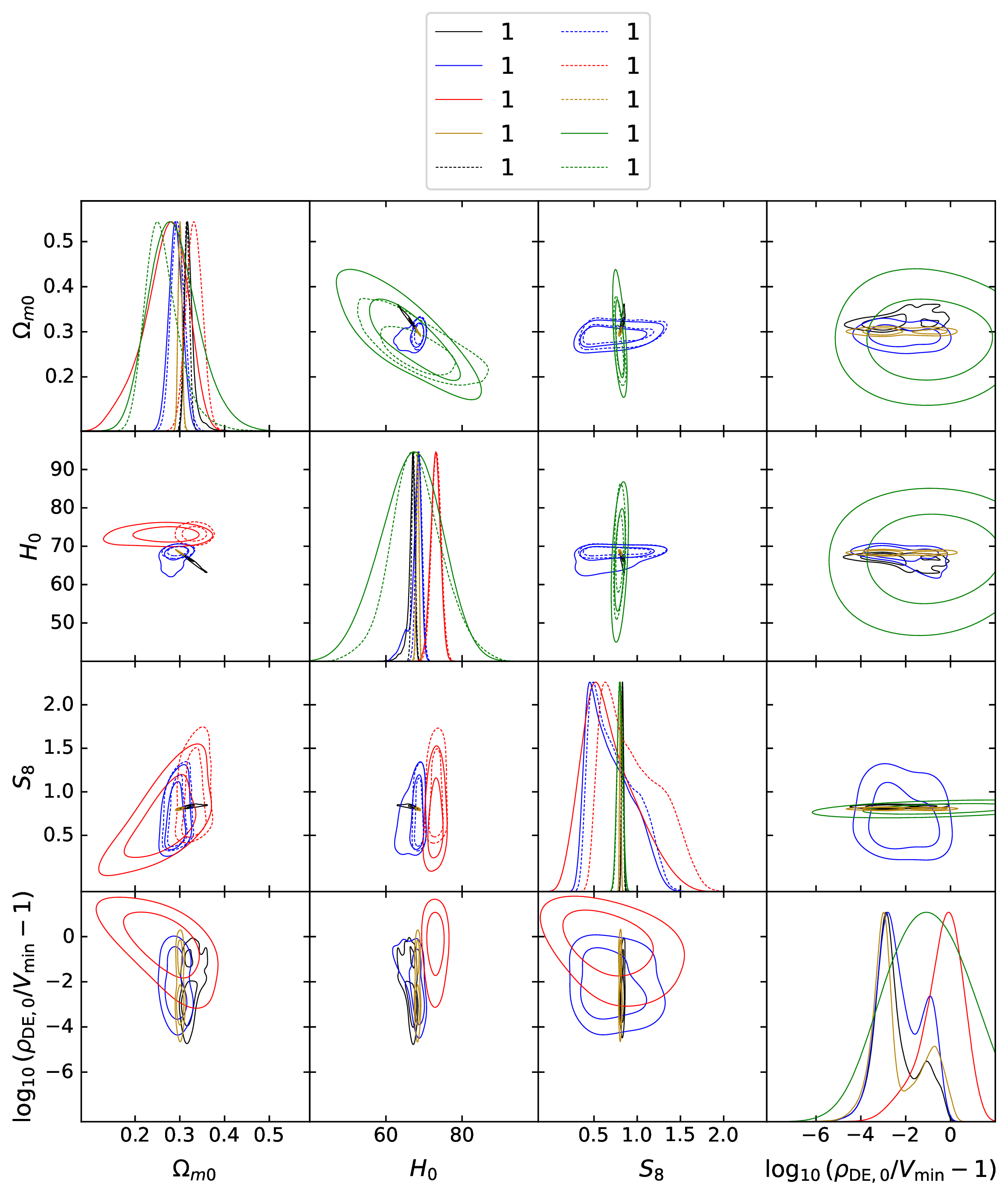}\\
\includegraphics[trim = 900 1900 0 0, clip, width = 0.4\textwidth]{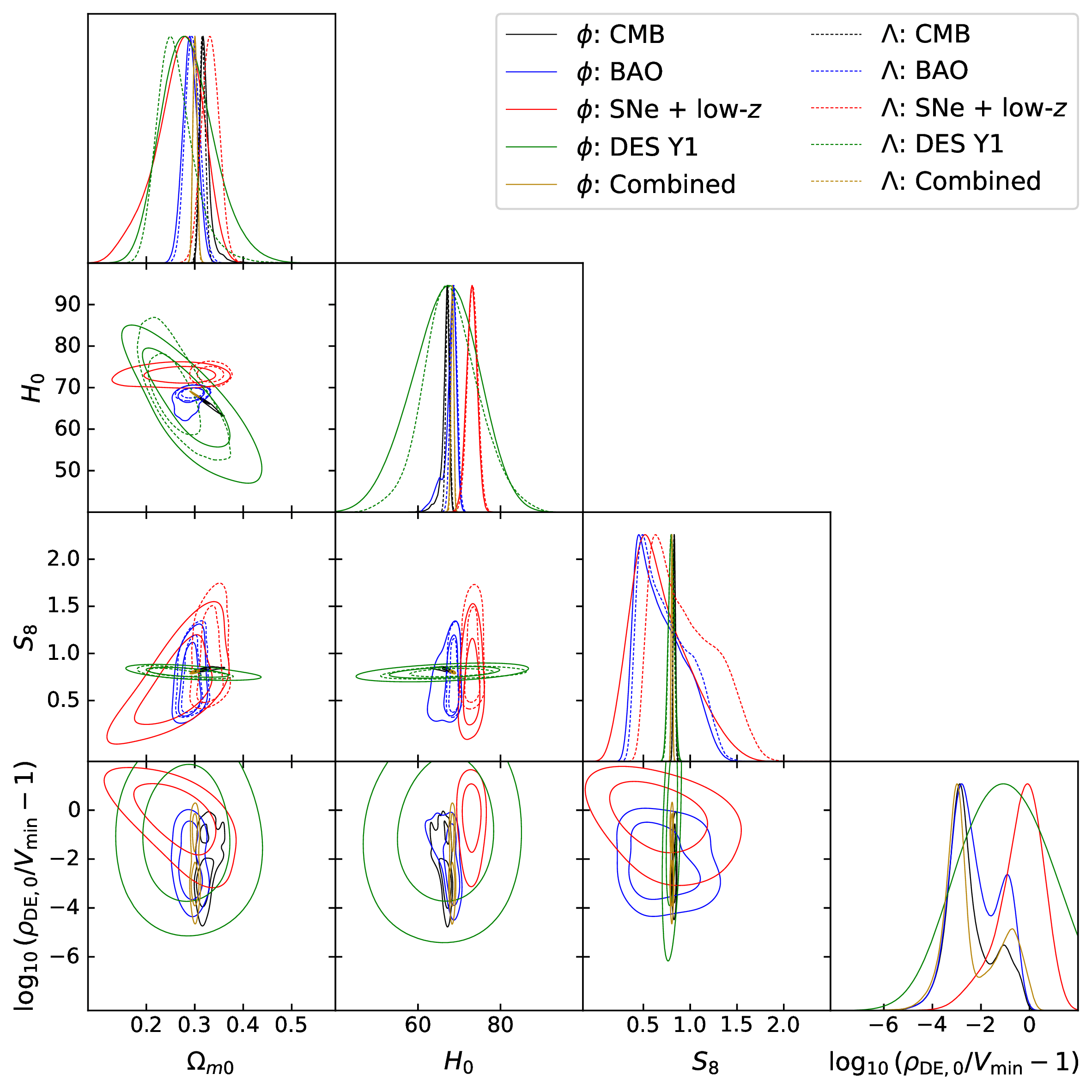}
\caption{\label{fig:fit_individual_late_params} 68\% and 95\% C.L. posterior distribution of the late-time quantities from CMB (Black), BAO ({\color{RoyalBlue}blue}), SNe + low-$z$ ({\color{BrickRed}red}), and DES Y1 ({\color{Brown}Brown}) data, with $\Lambda$CDM model the \COMRev{lighter shade}\Rev{dashed} contours and the axion-like dark energy model \COMRev{in section~\ref{sec:model}}\Rev{($\phi$CDM)} the \COMRev{darker shade}\Rev{solid} contours. \COMRev{The strong tension between astrophysical data of SNe + low-$z$ and cosmological data of CMB + BAO presented in $\Lambda$CDM model posterior of $H_0$ persists in the axion-like dark energy model. }The lack of constraint on $S_8$ for BAO and SNe + low-$z$ datasets is expected as we impose no a priori constraint on the primordial perturbation. Note that the DES Y1 posterior is a rough estimate, as explained in section~\ref{sec:data}.}
\end{figure}

\COMRev{However}\Rev{Yet}, as shown in Fig.~\ref{fig:fit_individual_late_params}, the $2$-dimensional posterior distribution shifts significantly from $\Lambda$CDM model to the axion-like dark energy model, especially for SNe + low-$z$ dataset. This may contribute to the very slight ease of tension between CMB + BAO dataset and SNe + low-$z$ dataset, as shown in table~\ref{table:tension_scalar}. \COMRev{It mainly comes from}\Rev{This is a direct consequence of} Eq.~\eqref{sfeosgammamodel}, where the increase of dark energy equation of state during the tracking regime leads to less dark matter at the current time than what would be expected from the luminosity distance measurements\Rev{, and is common among dark energy models with an equation of state deviating from
-1}. \COMRev{Unfortunately, the Hubble tension remains despite the introduction of the dynamical field.}

\begin{figure}
\centering
\includegraphics[width = 0.8 \textwidth]{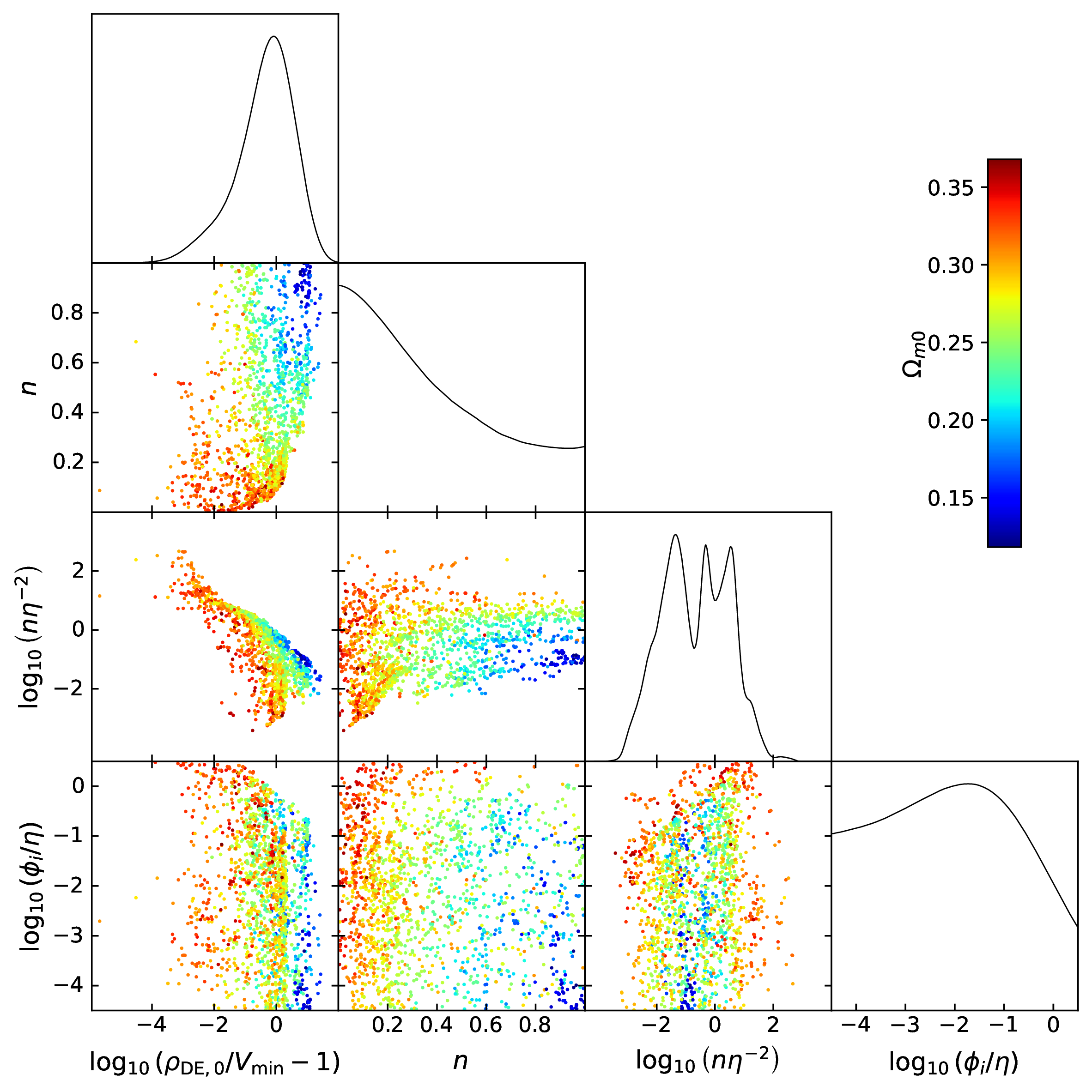}
\caption{\label{fig:fit_scatter_SN_LowZ} Scatter plot of the late-time parameters for the axion-like dark energy model in section~\ref{sec:model} of SNe + low-$z$ dataset, colour-coded by the current matter density $\Omega_{m0}$. All extended parameters have strong dependence with $\Omega_{m0}$, especially $n$. This is consistent with the prediction of Eq.~\eqref{sfeosgammamodel}.
}
\end{figure}

\begin{figure}
\centering
\includegraphics[width = 0.8 \textwidth]{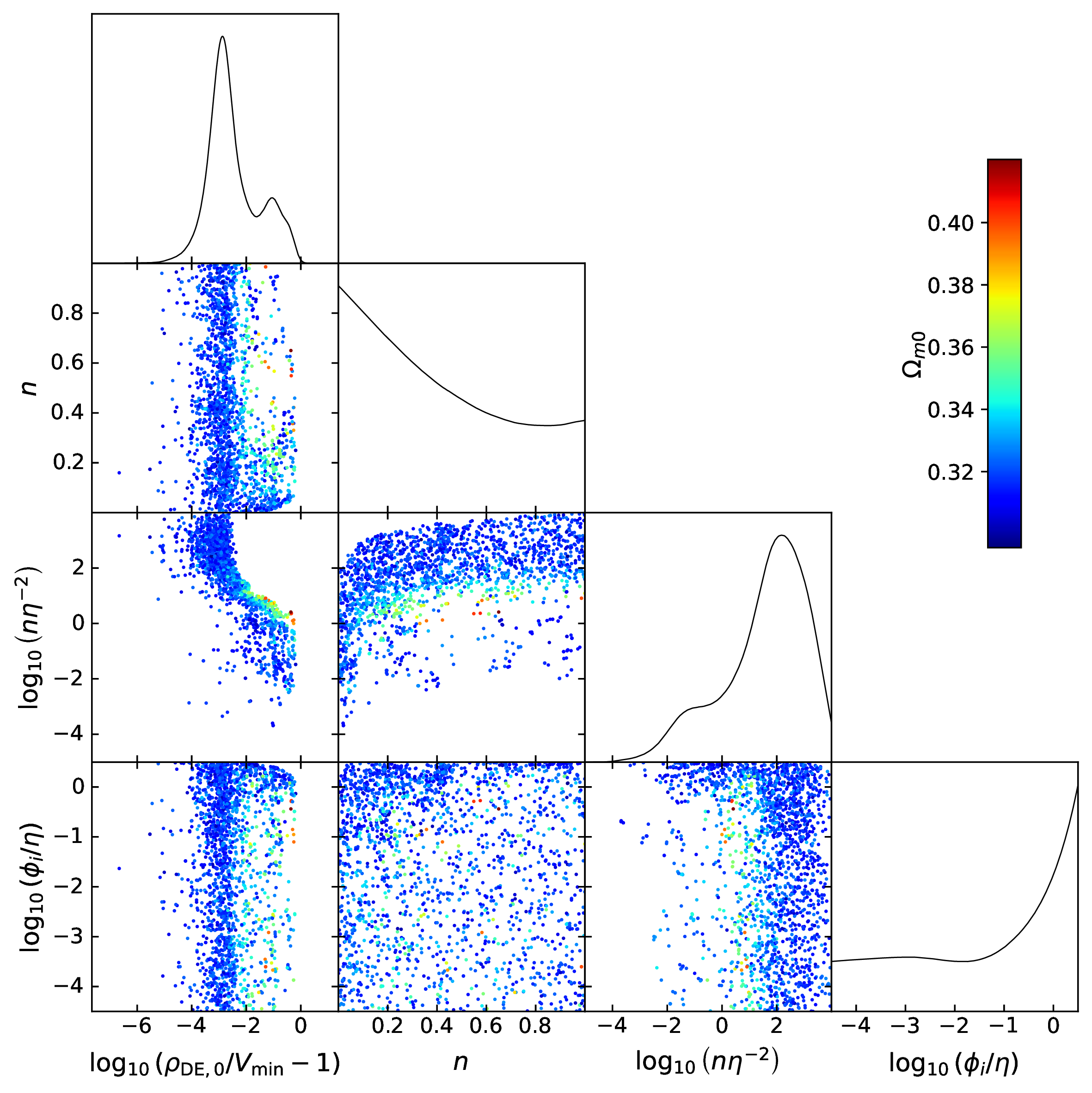}
\caption{\label{fig:fit_scatter_CMB} Scatter plot of the extended parameters for the axion-like dark energy model in section~\ref{sec:model} of CMB dataset, colour-coded by $\Omega_{m0}$. If the scalar field is in the tracking regime long enough ($\log(n\eta^{-2}) < 2$ and $\log(\phi_i/\eta) < 0$), the equation of state $w_\phi$ would be greater than $-1$ for a period at late-time, leading to longer distance to the last scattering surface than $\Lambda$CDM prediction, and thus lower $H_0$ and greater $\Omega_{m0}$ to compensate. The bimodal behaviour of the excess dark energy density $\log_{10}(\rho_{\rm DE,0}/V_{\rm min} - 1)$ is an artifact of the Gaussian constraint on $\mathcal{E}_{H_0}$ in table~\ref{table:prior}, and would not affect our conclusion that the axion-like dark energy model cannot resolve the Hubble tension.
}
\end{figure}

To better understand how the axion-like dark energy model affects the cosmological fit\COMRev{ and why it cannot solve the Hubble tension}, the extended parameters of the axion-like dark energy model are shown in Fig.~\ref{fig:fit_scatter_SN_LowZ} for SNe + low-$z$ dataset and Fig.~\ref{fig:fit_scatter_CMB} for CMB dataset as scatter plots over the current matter density parameter $\Omega_{m0}$. We choose to omit $H_0$ as it is either fully constrained (SNe + low-$z$) or is strongly correlated with $\Omega_{m0}$ (CMB) and thus provides no additional information.

For Fig.~\ref{fig:fit_scatter_SN_LowZ}, the effect of the axion-like dark energy on $\Omega_{m0}$ prediction of SNe + low-$z$ dataset is tremendous. \COMRev{T}\Rev{Again, t}his is the direct consequence of Eq.~\eqref{sfeosgammamodel} that a larger scalar equation of state $w_\phi$ leads to a lower value of $\Omega_{m0}$ for a constant $H_0$. Furthermore, one can clearly identify the tracking regime in $\log(n \eta^{-2})$ - $\log(\phi_i/\eta)$ scatter plot. Heavier fields ($\log(2m^2/V_{\rm min})=\log(n \eta^{-2}) \gg 0$) roll down the potential faster, thus ending the tracking regime at higher redshift, affecting SNe + low-$z$ dataset prediction less. Lighter fields ($\log(n \eta^{-2}) \ll 0$) on the other hand seems to be clustered around $n\sim0$, suggesting that they are effectively $w$CDM with $w\sim-1$. To have non-trivial effect on SNe + low-$z$ prediction, the scalar field should exit the tracking regime during dark energy dominant era ($\log(n \eta^{-2}) \sim 0$ and $\log(\phi_i/\eta) < 0$) and should have distinguishable feature during the tracking regime ($n \nsim 0$).

For Fig.~\ref{fig:fit_scatter_CMB} though, the effect of the axion-like dark energy on $\Omega_{m0}$ prediction of CMB dataset is less pronounced. $\Omega_{m0}$ is pulled toward the opposite direction than the prediction of SNe + low-$z$ dataset. This is, again, consistent with Eq.~\eqref{sfeosgammamodel} that a larger scalar equation of state $w_\phi$ (greater $H(z)/H_0$) necessitates a lower value of $H_0$ (effectively lower $\Omega_{m0}$ for a constant $\Omega_{m0} h^2$) to maintain a constant distance to the last scattering surface. The deviation from $w = -1$ is more effective at lower redshift as it has greater effect on the distance measure. A lighter field ($\log(n\eta^{-2}) < 2$) that stays long enough inside the tracking regime ($\log(\phi_i/\eta) < 0$) thus has greater effect on $\Omega_{m0}$. Unfortunately, Fig.~\ref{fig:fit_scatter_CMB} is polluted by the finite width Gaussian constraint of $\mathcal{E}_{H_0}$ in table~\ref{table:prior}. The finite width causes an oversampling of heavier fields that have reached the bottom (the Gaussian peak at $\log_{10}(\rho_{\rm DE,0}/V_{\rm min} - 1) \sim -3$), but would not affect the conclusion that the axion-like dark energy \COMRev{in section~\ref{sec:model} }predicts a lower value of $H_0$ for CMB dataset\COMRev{, thus unable to ease the Hubble tension}\Rev{ than the $\Lambda$CDM prediction if the tracking regime overlaps with the dark-energy-dominated era. The inclusion of the low-$z$ dataset eliminates this possibility, forcing the axion-like dark energy model to follow closely with $\Lambda$CDM. The oversampling from the finite width of $\mathcal{E}_{H_0}$ thus merely alters the posterior of the extended parameters and would not affect the result. For a more detailed analysis, see appendix~\ref{sec:prior}}.

Combining two datasets together suggests that the preferred configuration of the axion-like dark energy model is either extremely heavy ($\log(n\eta^{-2}) > 2$) such that the system reduces back to effectively a $\Lambda$CDM model, or lighter ($\log(n\eta^{-2}) < 0$) and with small enough initial field strength ($\log(\phi_i/\eta) \gtrsim 0$) that we are just entering the tracking regime. The later solves the weak $\Omega_{m0}$ tension between CMB + BAO and SNe + low-$z$ datasets ($1.4\sigma$ 
in $\Lambda$CDM model to $0.6\sigma$ 
in the axion-like scalar model), and is reflected by the slight decrease of suspiciousness and the goodness of fit test as shown in table~\ref{table:tension_scalar}. This bimodal behaviour can be seen in Fig.~\ref{fig:fit_individual_late_params}, with the heavier branch reaching the bottom ($\log(\rho_{\rm DE,0}/V_{\rm min} -1) \ll 0$) and the lighter branch still freezes on the potential slope ($\log(\rho_{\rm DE,0}/V_{\rm min} -1) \gtrsim 0$).

\subsection{Cosmic evolution, CMB power spectrum, matter power spectrum and $f\sigma_8$}
\label{sec:evolution}

\begin{figure}
\centering
\subfigure[CMB: $w_\phi (a/a_0)$\label{fig:w_CMB}]{\includegraphics[trim = 40 0 175 135, clip, width = 0.49 \textwidth]{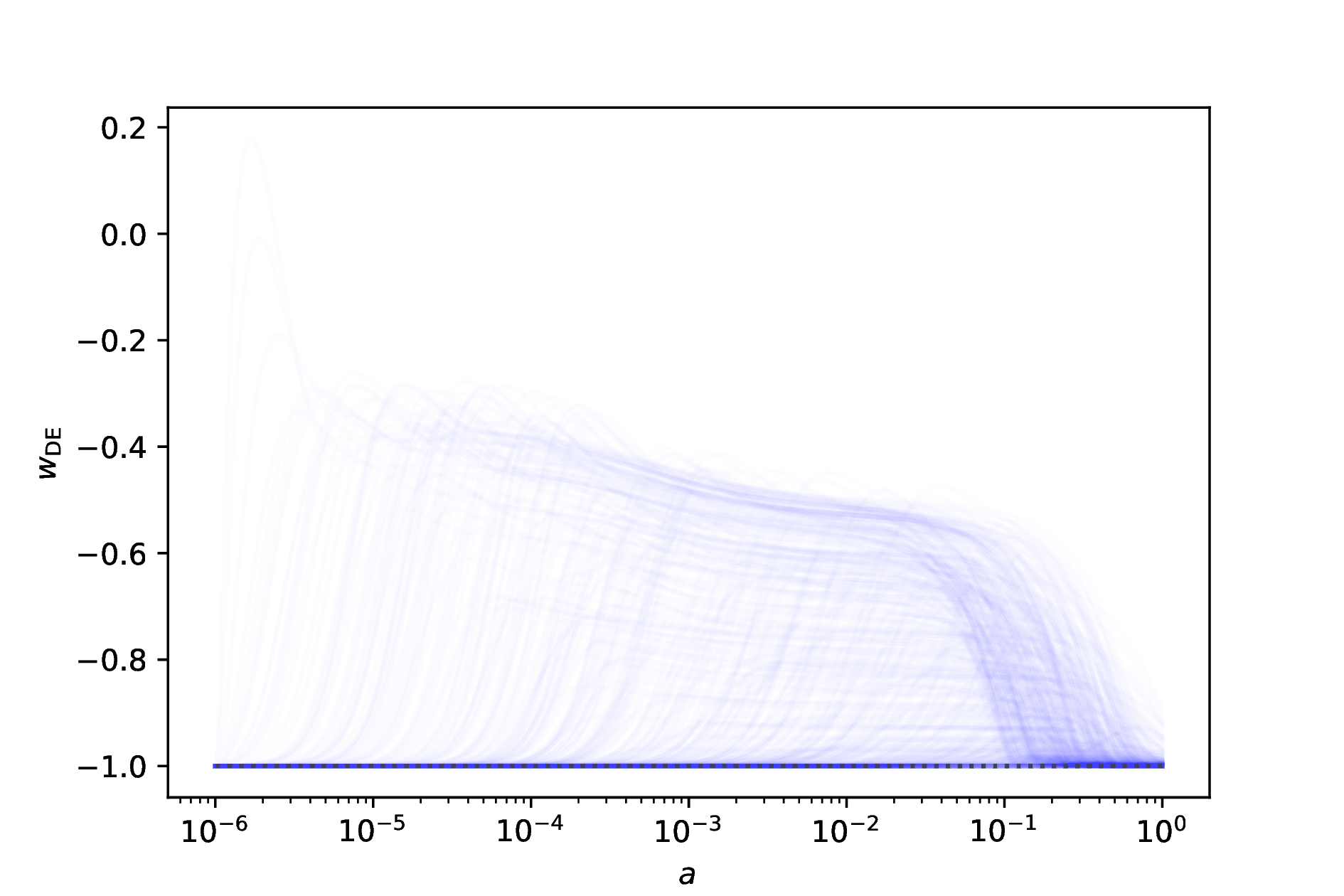}}~
\subfigure[CMB + BAO: $w_\phi (a/a_0)$\label{fig:w_CMB_BAO}]{\includegraphics[trim = 30 0 185 135, clip, width = 0.49 \textwidth]{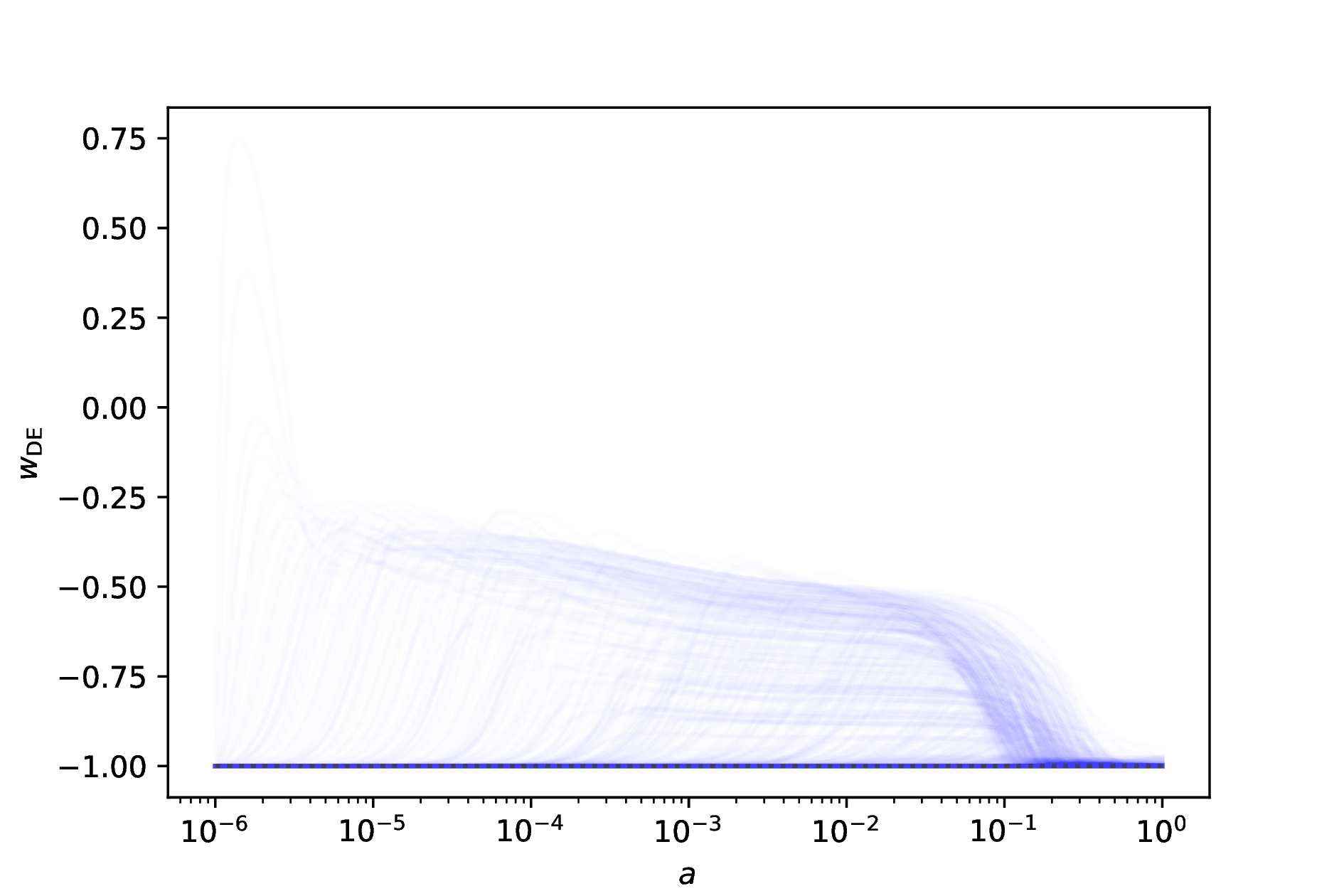}}
\\
\subfigure[CMB: $\Omega_r$ ({\color{RoyalBlue}blue}), $\Omega_n$ ({\color{Dandelion}yellow}), $\Omega_\phi$ ({\color{Purple}purple})\label{fig:Omega_CMB}]{\includegraphics[trim = 60 0 155 115, clip, width = 0.49 \textwidth]{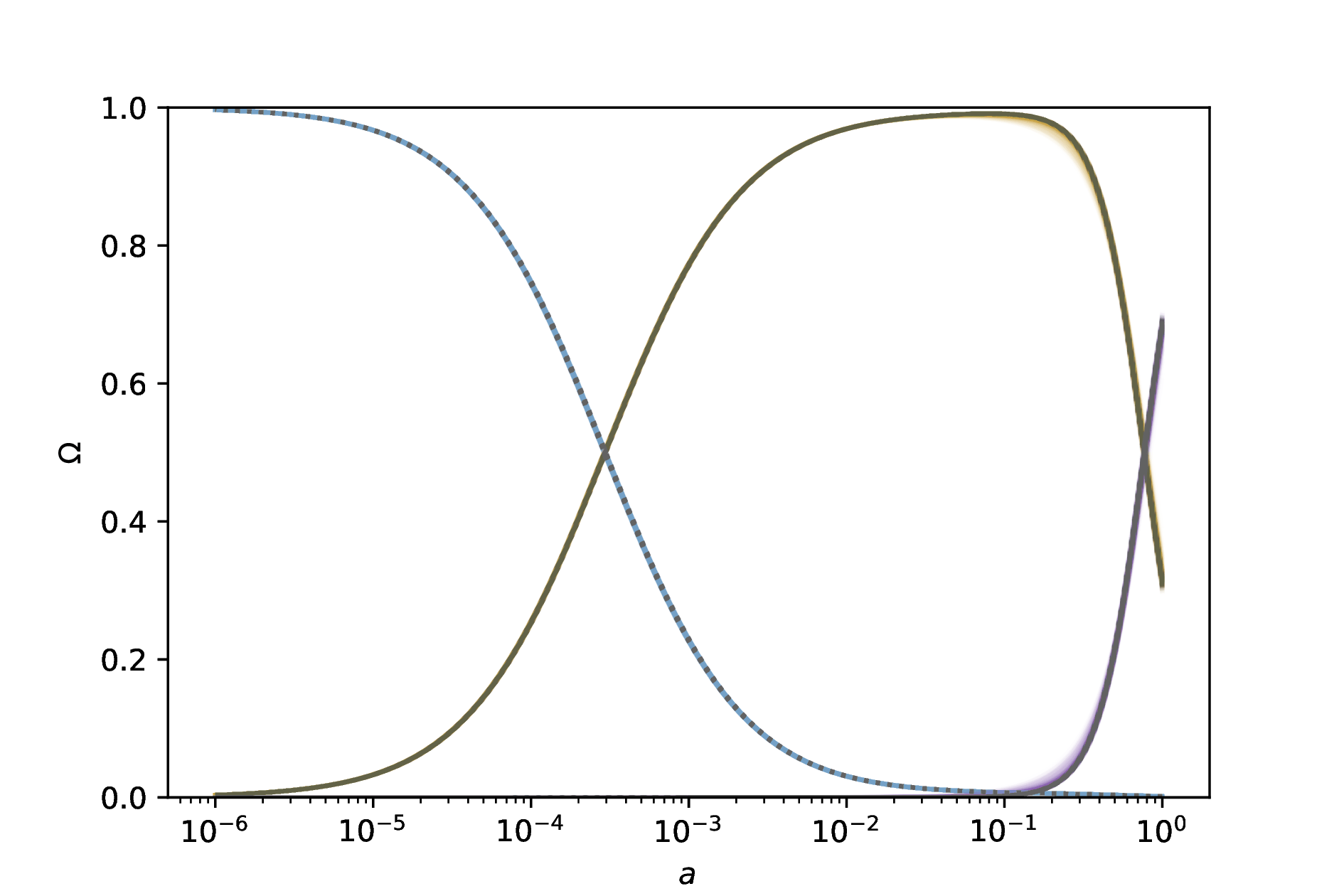}}~
\subfigure[CMB + BAO: $\Omega_r$ ({\color{RoyalBlue}blue}), $\Omega_m$ ({\color{Dandelion}yellow}), $\Omega_\phi$ ({\color{Purple}purple})\label{fig:Omega_CMB_BAO}]{\includegraphics[trim = 55 0 160 115, clip, width = 0.49 \textwidth]{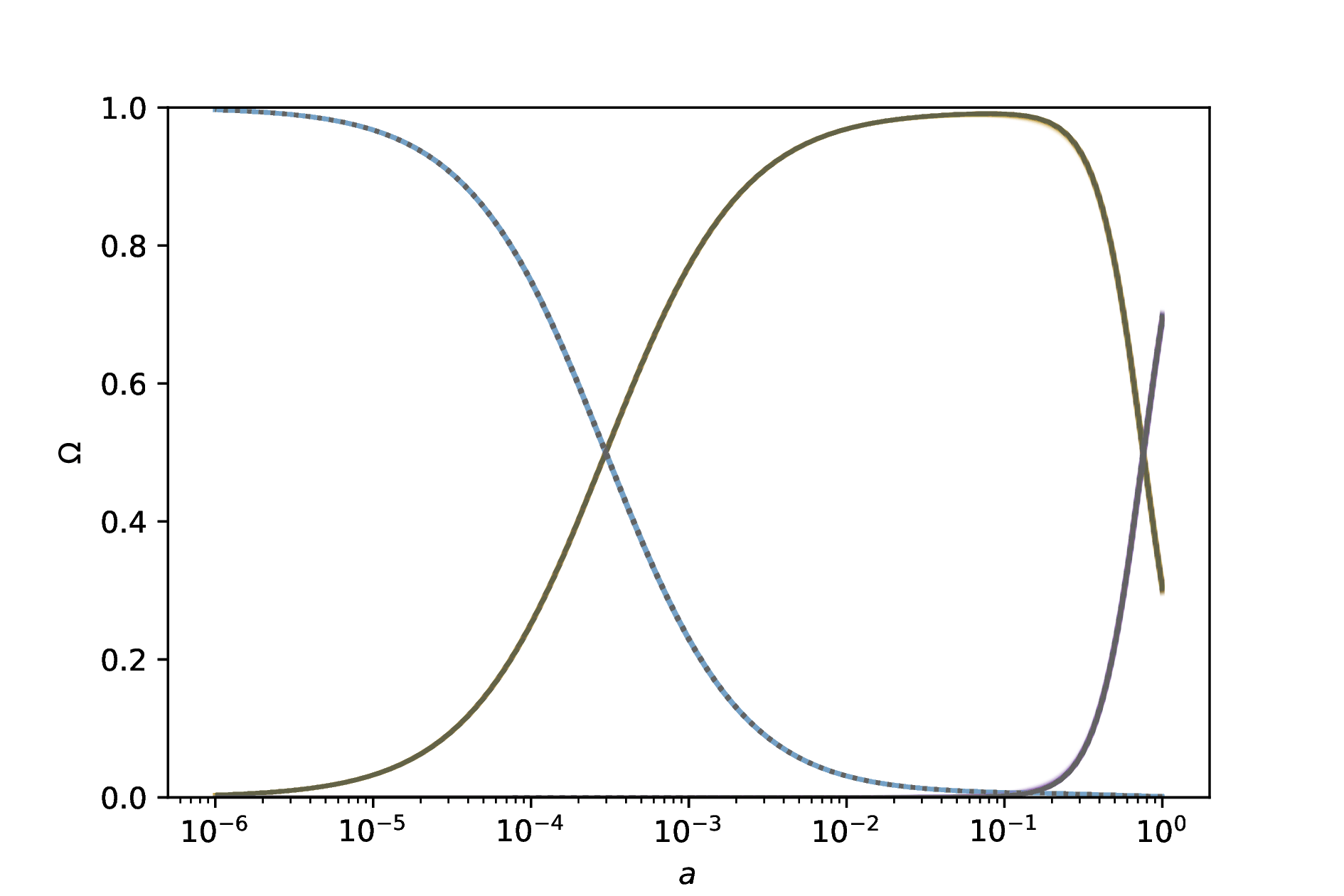}}
\\
\subfigure[CMB: $f \sigma_8$\label{fig:fsigma8_CMB}]{\includegraphics[width = 0.43 \textwidth]{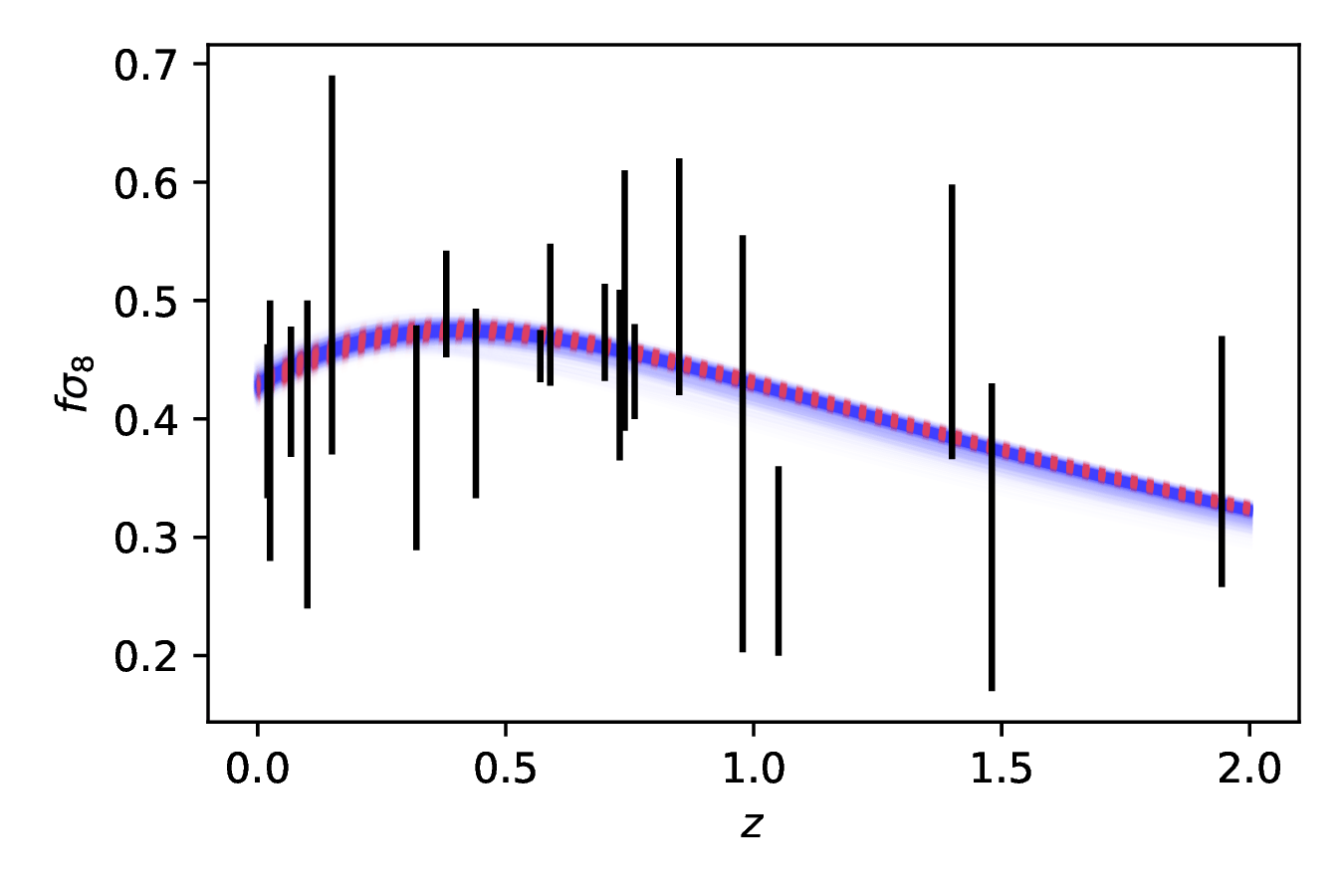}}~
\subfigure[CMB + BAO: $f \sigma_8$\label{fig:fsigma8_CMB_BAO}]{\includegraphics[width = 0.43 \textwidth]{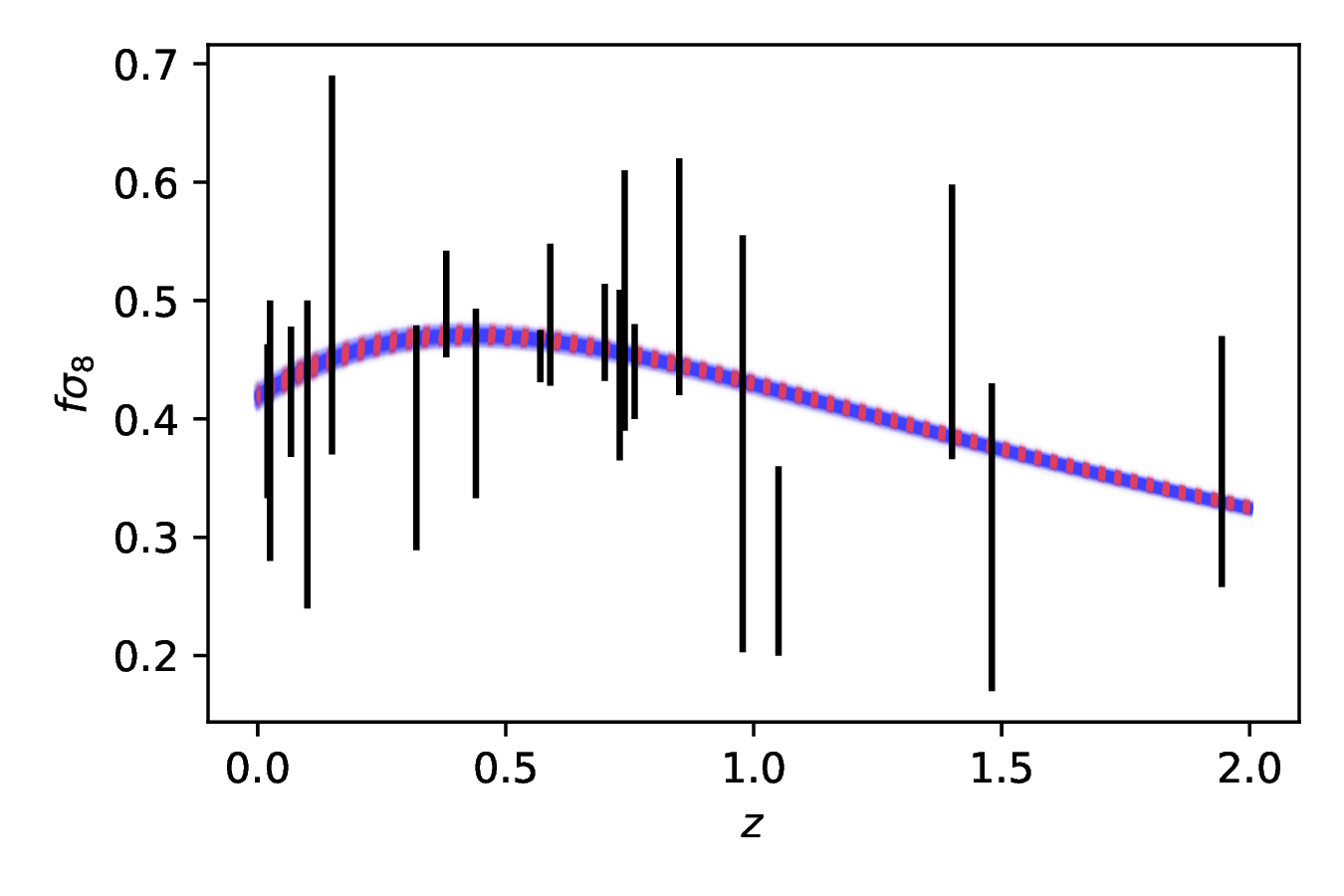}}
\caption{\label{fig:cosmic_evolution}
The dark energy equation of state $w_\phi$ (\emph{top}), density parameters (\emph{centre}) of radiation $\Omega_r$ ({\color{RoyalBlue}blue}), matter $\Omega_m$ ({\color{Dandelion}yellow}) and dark energy $\Omega_\phi$ ({\color{Purple}purple}) as functions of $a/a_0$, and $f\sigma_8$ (\emph{bottom}) as functions of $z$ for 600 samples drawn from individual fits. \emph{Left} figures: Models fitted against CMB dataset. \emph{Right} figures: Models fitted against CMB + BAO dataset. \emph{Black} line denotes $\Lambda$CDM model and non-black lines denote the axion-like dark energy model. Without additional distance measurement, CMB dataset alone permits variation of the density parameter evolution as depicted in Fig.~\ref{fig:omegasevolution}. However, once BAO dataset is included, the density parameters stabilise and the tracking regime has to end before $a/a_0 = 1$. In \emph{bottom} figures, the $f\sigma_8$ data are taken from table~2 of \cite{Avila:2022xad}.
}
\end{figure}

As shown in Fig.~\ref{fig:omegasevolution}, the effect of the axion-like dark energy model on the cosmic evolution could be substantial. We would like to know if such effect persists when we impose observational constraint. In Fig.~\ref{fig:cosmic_evolution}, we show explicitly that CMB dataset alone permits modification to the density parameter evolution and the growth history, with the scalar field still in the tracking regime today. However, once the BAO dataset is included, the additional distance measure by BAO observations stabilises the density parameters by ending the tracking regime early. The resulting density evolution and the growth history closely follow the evolution of $\Lambda$CDM model, despite the existence of the tracking regime right before the current era\footnote{One should be cautious to compare Figs.~\ref{fig:fsigma8_CMB} and \ref{fig:fsigma8_CMB_BAO} with Fig.~5(b) of \cite{Boiza:2024fmr} and Fig.~5 of \cite{Hossain:2025grx}, as what we present here is derived from the cosmological fit instead of an a priori parameter set.}. This poses an interesting question of whether the tracking behaviour would modify the power spectrum at a certain scale.

\begin{figure}
\centering
\begin{tikzpicture}[      
        every node/.style={anchor=south west,inner sep=0pt},
        x=1mm, y=1mm,
      ]   
     \node (fig2) at (0,0)
       {\includegraphics[width = \textwidth]{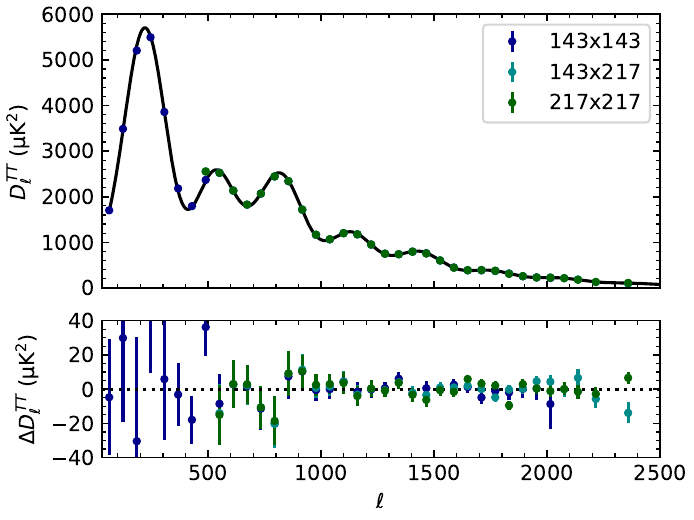}};  
     \node (fig1) at (2.7,9.1)
       {\includegraphics[width = 1.04\textwidth]{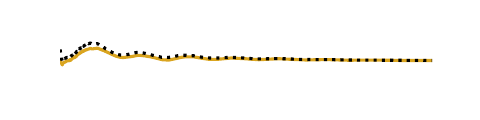}};
     \node (fig3) at (2.7,9.1)
       {\includegraphics[width = 1.04\textwidth]{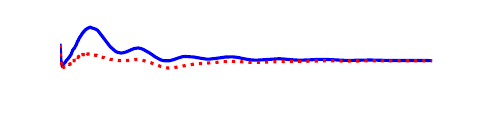}};
\end{tikzpicture}
\caption{\label{fig:power_spectrum_high_l}
\emph{Top}: The best fit TT power spectrum of $\Lambda$CDM model against Planck CamSpec PR4 high-$\ell$ TTTEEE data release and the PR4 high-$\ell$ TT data points in Fig.~6 of \cite{Rosenberg:2022sdy}. \emph{Bottom}: The residue of the best fit TT power spectrum and the PR4 high-$\ell$ TT data points against best fit $\Lambda$CDM model of Planck CamSpec PR4 high-$\ell$ TTTEEE dataset. Black \emph{dotted} line and {\color{BrickRed}red} \emph{dotted} line denote $\Lambda$CDM model fitted against CMB and CMB + BAO datasets respectively, and {\color{Dandelion}yellow} \emph{solid} line and {\color{RoyalBlue}blue} \emph{solid} line denotes axion-like dark energy model fitted against CMB and CMB + BAO datasets respectively. There are differences between two models but it is hard to tell the statistical significance. The difference between the CMB-only best fit $\Lambda$CDM model in our analysis and the Planck CamSpec PR4 high-$\ell$ TTTEEE best fit model in \cite{Rosenberg:2022sdy} stems from the difference in data inclusion, as we also include low-$\ell$ TTEE dataset \cite{Planck:2019nip} and PR4 CMB lensing dataset \cite{Carron:2022eyg, Carron:2022eum}. 
}
\end{figure}

\begin{figure}
\centering
\begin{tikzpicture}[      
        every node/.style={anchor= west,inner sep=0pt},
        x=\textwidth, y=\textwidth,
      ]   
     \node (fig2) at (0,0)
       {\includegraphics[width = \textwidth]{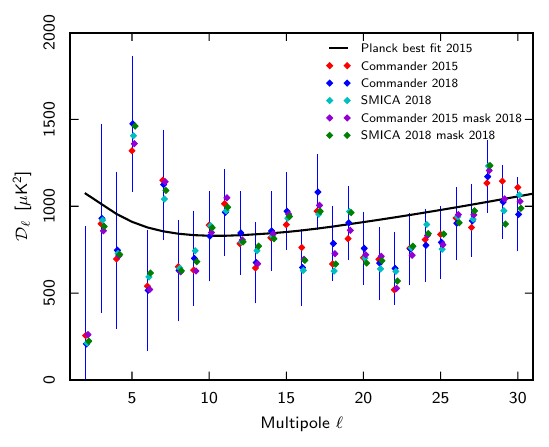}};
     \node (fig1) at (-0.008,0.035)
       {\includegraphics[width = 1.09\textwidth]{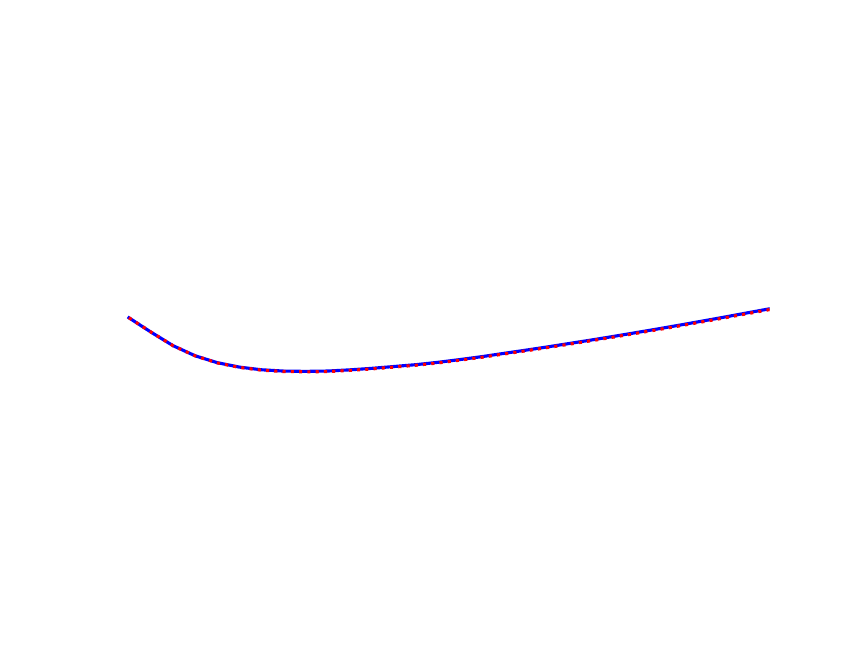}};
     \node (fig1) at (-0.008,0.035)
       {\includegraphics[width = 1.09\textwidth]{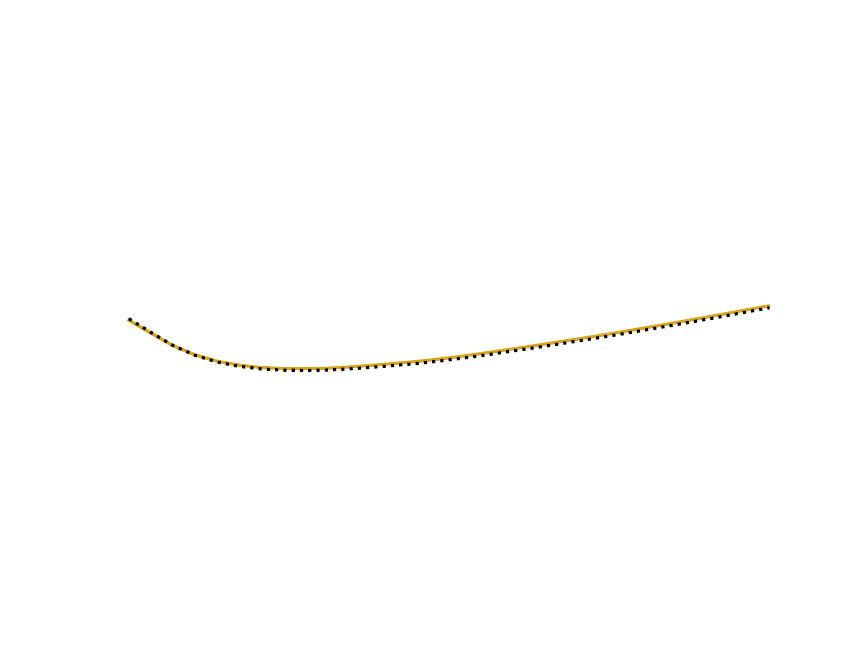}};
\end{tikzpicture}
\caption{\label{fig:power_spectrum_lowl}
The best fit CMB TT low-$\ell$ power spectra (\emph{top}) and the data of various maps and masks from Planck18 data release in Fig.~2 of \cite{Planck:2019nip}. Black solid line is the best fit $\Lambda$CDM model prediction from Planck15. Black \emph{dotted} line and {\color{BrickRed}red} \emph{dotted} line denote $\Lambda$CDM model fitted against CMB and CMB + BAO datasets respectively, and {\color{Dandelion}yellow} \emph{solid} line and {\color{RoyalBlue}blue} \emph{solid} line denotes axion-like dark energy model fitted against CMB and CMB + BAO datasets respectively. There are no visible difference across all cases except the best fit model of Planck15.
}
\end{figure}

\begin{figure}
\begin{tikzpicture}[      
        every node/.style={anchor=south west,inner sep=0pt},
        x=1mm, y=1mm,
      ]   
     \node (fig2) at (0,0)
       {\includegraphics[width = \textwidth]{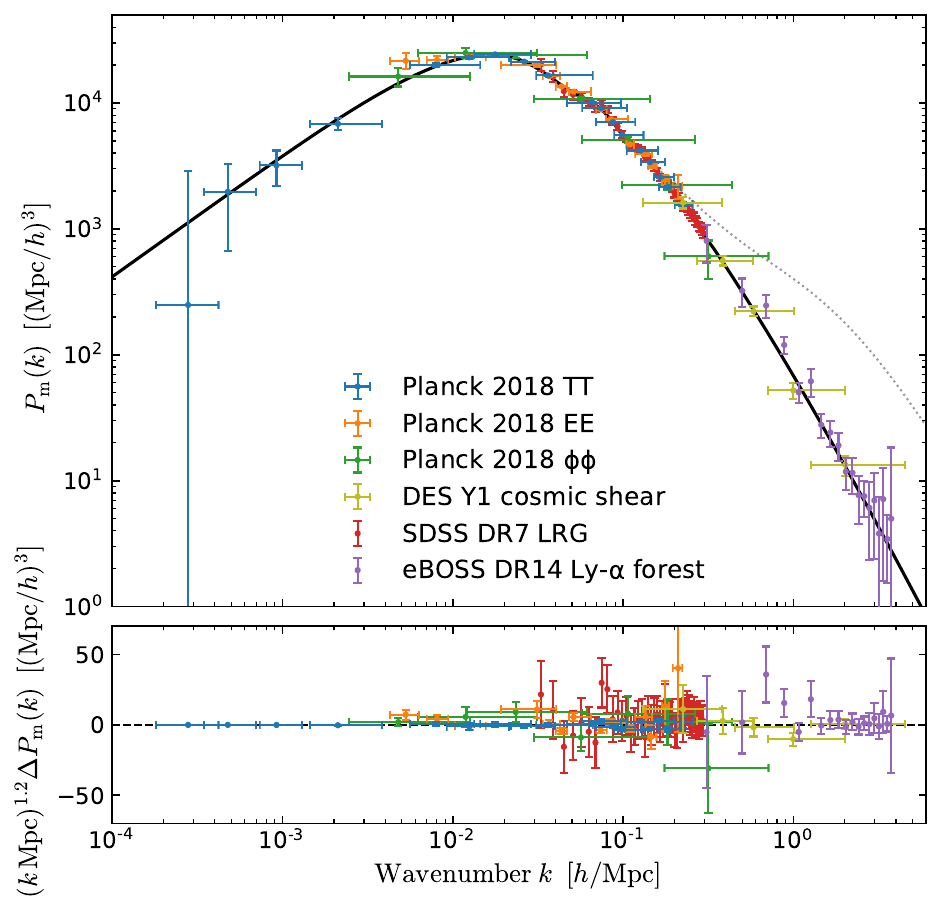}};
     \node (fig1) at (-2.8,10.7)
       {\includegraphics[width = 1.115\textwidth]{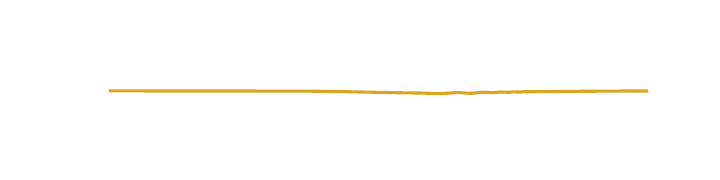}};
     \node (fig3) at (-2.8,10.7)
       {\includegraphics[width = 1.115\textwidth]{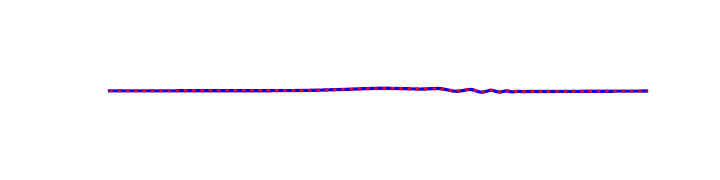}};
\end{tikzpicture}
\caption{\label{fig:matter_power_spectrum}
The best fit matter power spectrum (\emph{top}) and the residue (\emph{bottom}) against best fit $\Lambda$CDM model matter power spectrum of CMB dataset. \emph{Top}: \emph{Solid} lines represents the linear power spectrum and the \emph{dotted} lines represents the nonlinear power spectrum. Matter power spectrum data taken from Fig.~1 of \cite{Chabanier:2019eai}. \emph{Bottom}: Black \emph{dashed} line and {\color{BrickRed}red} \emph{dotted} line denote $\Lambda$CDM model fitted against CMB and CMB + BAO datasets respectively, and {\color{Dandelion}yellow} \emph{solid} line and {\color{RoyalBlue}blue} \emph{solid} line denotes axion-like dark energy model fitted against CMB and CMB + BAO datasets respectively. The axion-like dark energy model shows slight suppression at acoustic scale for CMB only dataset, but two models line up once BAO data is included.
}
\end{figure}

To better understand the effect of the tracking behaviour, we isolate the maximum-a-posteriori prediction of the axion-like dark energy model and the $\Lambda$CDM model by the datasets of CMB and CMB + BAO and compare them in detail against the prediction of best fit $\Lambda$CDM model from Planck18. As shown in Figs.~\ref{fig:power_spectrum_high_l}, \ref{fig:power_spectrum_lowl} and Fig.~\ref{fig:matter_power_spectrum}, there are no clear difference between models from the perspective of CMB power spectrum\footnote{The wild oscillation of the CMB power spectrum at the acoustic peak scale is expected from the variation of the distance to the last scattering surface, induced by the difference in the late-time cosmic evolution. This oscillation is statistically insignificant.} and matter power spectrum once including the BAO data. It appears that despite the existence of the tracking behaviour up to relatively recent era, the axion-like dark energy model cannot affect either the CMB power spectrum or the matter power spectrum. 

\section{Conclusions}
\label{sec:conclusion}

In this work, we have first revisited the generalised axion-like dark energy model, which extends axion-like potentials to negative exponents. We have examined its key dynamical properties, showing how it accounts for late-time cosmic acceleration through an effective cosmological constant at the minimum of the potential. The model exhibits tracking behaviour, helping to alleviate the coincidence problem, and naturally enhances the transition from matter domination to dark energy. For more detailed discussions on the theoretical aspects of the model, we refer the reader to \cite{Hossain:2023lxs,Boiza:2024azh,Boiza:2024fmr,Hossain:2025grx}.

We have fitted the axion-inspired scalar field dark energy model with a potential of the inverse power of cosine in Eq.~\eqref{axionscalarpotential} from \cite{Boiza:2024azh} using the complete cosmological datasets of CMB (Planck2018 low-$\ell$ TTEE \cite{Planck:2019nip}, NPIPE CamSpec high-$\ell$ TTTEEE \cite{Rosenberg:2022sdy}, PR4 CMB lensing \cite{Carron:2022eyg, Carron:2022eum}) , BAO (DESI DR1 without full-shape \cite{DESI:2024mwx, DESI:2024lzq, DESI:2024uvr}), supernovae (Pantheon+ SNe catalog \cite{Brout:2022vxf} without anchoring of SNe standardised absolute magnitude), low-$z$ $H_0$ measurements (Riess et al. low-$z$ anchors \cite{Riess:2020fzl}), and DES Y1 morphology \cite{DES:2017myr}. In addition to the existence of the tracking solution common to many scalar field dark energy models that solves the initial condition problem, the model considered in this work aims to solve the coincidence problem of why the dark energy only dominates at $z\lesssim1$ with $w\sim-1$ by offering a graceful exit of the tracking regime. \COMRev{The model seems promising from the perspective of naturalness in parameter selection, as the system closely follows the $\Lambda$CDM model for the majority of the time, except during the matter dominated era.}\Rev{The main attraction of the model is its insensitivity to the model parameters, as the system closely follows the $\Lambda$CDM model, except for a finite duration after matter-radiation crossover.} We hoped that the model may ease the $S_8$ tension without affecting the Hubble tension through an early matter-dark energy transition, as demonstrated in \cite{Boiza:2024fmr}.

\COMRev{Unfortunately, t}\Rev{T}he model shows no statistical\COMRev{ly meaningful benefit over $\Lambda$CDM model across Planck, DESI, Pantheon+, Riess et al. low-$z$ anchors and DES Y1 morphological datasets. A reduction in tension by $0.1\pm0.3\sigma$ is realised between high-$z$ dataset of Planck and DESI, and low-$z$ dataset of Pantheon+ and $H_0$ anchors, at the cost of an increase in tension by $0.3\pm0.5\sigma$ between the combined dataset of Planck, DESI, Pantheon+, $H_0$ anchors, and the low-$z$ matter perturbation dataset of DES Y1. This clearly cannot solve the cosmological tension. In addition}\Rev{   disadvantage nor meaningful advantage over $\Lambda$CDM model across datasets considered in this work. A reduction in tension by $0.1\pm0.3\sigma$ is realised between high-$z$ dataset of Planck and DESI, and low-$z$ dataset of Pantheon+ and $H_0$ anchors, but is compensated by an increase of $0.3\pm0.5\sigma$ in the tension with the low-$z$ weak lensing dataset of DES Y1. In particular}, the structure formation history remains practically identical to the one in $\Lambda$CDM model. We isolate the root cause to the inability of the axion-like dark energy model to increase $H_0$, and thus decrease $\Omega_{m0}$ under the constraint of Planck CMB data. The salient feature of the model, a long-lasting $w>-1$ tracking regime during the matter-dominated era with a graceful exit, is thus rendered ineffective.

Despite the similarity with the $\Lambda$CDM model from the observation point of view, the axion-like scalar dark energy model does provide interesting features and dynamics that distinguish it from either a cosmological constant or other quintessence models. The existence of the tracking region implies that the observationally preferred field is either so heavy that it has reached the bottom, or so light that it has not started rolling. When incorporating the axion-like dark energy model with non-trivial dark matter coupling, early dark energy, phantom, etc. through multi-fields, this bimodal behaviour may serve as an additional constraint on the model space.

In conclusion, while the axion-like scalar dark energy model introduced in \cite{Boiza:2024azh} remains competitive with $\Lambda$CDM model across a wide area of the parameter space, it lacks obvious observational features to separate one from another either. Meanwhile, the model provides interesting features such as a tracking regime with graceful exit that distinguish it from either a cosmological constant or other quintessence models from the theoretical point of view. Therefore, it serves as a reasonable vehicle for future extension.

\appendix
\section{Statistical probes}
\label{sec:stat}

The definition of each probe is listed below:
\begin{align}
\ln B (D|M) &\equiv \ln \int P(D|\theta_M) P(\theta_M) d\theta_M = -\ln V_M - \ln \left< \left( P(D|\theta_M) \right)^{-1} \right>_{M|D}  \label{eq:BE}  \,,\\
{\rm DIC} (D|M) &\equiv \ln P(D|\theta_{M,\,map}) - 2F(D|M)  \,,\\
{\rm WAIC} (D|M) &\equiv - F(D|M) + {\rm BMD} (D|M) / 2  \,,\\
F(D|M) &\equiv \left< \ln P(D|\theta_M) \right>_{M|D} \equiv \ln B (D|M) + {\rm KL} (D|M)  \,,\\
{\rm BMD} (D|M) &\equiv 2 \left( \left< \left( \ln P(D|\theta_M) \right)^2 \right>_{M|D} - \left< \ln P(D|\theta_M) \right>_{M|D}^2 \right)  \,,\\
-\ln R (D_1, D_2 | M) &\equiv - \ln B (D_1 D_2 | M) + \ln B (D_1 | M) + \ln B (D_2 | M)  \,,\\
{\rm GoF} (D_1, D_2 | M) &\equiv - \ln P(D_1 D_2 |\theta_{M,\,map}) + \ln P(D_1 |\theta_{M,\,map}) + \ln P(D_2 |\theta_{M,\,map})  \,,\\
S (D_1, D_2 | M) &\equiv - F(D_1 D_2 | M) + F (D_1 | M) + F (D_2 | M)  \,,
\end{align}
where $D$ and $M$ denote the dataset and the model, $\theta_M$ are the model parameters, $P(D|\theta_M)$, $P(\theta_M)$, $P(\theta_M|D) \equiv P(D|\theta_M) P(\theta_M) / B(D|M)$ are the likelihood, prior probability and posterior probability, $\left< ... \right>_{M|D} = \int ... P(\theta_M|D) d\theta_M$ is the MCMC mean, $V_M \equiv \int P(\theta_M) d\theta_M$ is the prior volume, $F$ is the deviance, ${\rm KL} (D|M) = \left< \ln P(\theta_M|D) - \ln P(\theta_M) \right>_{M|D}$ is the Kullback-Leibler divergence, BMD is the abbreviation of Bayesian model dimension, and $map$ is the abbreviation of maximum-a-posteriori, i.e., the Bayesian estimate of the model parameters.

While probes such as DIC, WAIC and Bayesian ratio can be easily interpreted using Jeffreys’ scale as shown in table~\ref{table:jeffreys}
, the goodness of fit and suspiciousness require further processing. As these two probes follow the BMD-dimensional $\chi^2$ distribution \cite{PhysRevD.100.023512}, we may convert a value $p$ into traditional $\sigma$ value as ${\rm CDF}_1^{-1} \left( {\rm CDF}_{\rm BMD} \left( \sqrt{{\rm BMD} + 2p} \right) \right)$ where ${\rm CDF}_d (p) = \int_0^p e^{-\chi^2}\chi^{2d - 2}d\chi^2$ is the cumulative distribution function of a $d$-dimensional Gaussian distribution.

\begin{table}[]
\centering
\begin{tabular}{|c|c|}
\hline
$\ln B (M_2) - \ln B (M_1)$, $\ln R$, etc.    & Interpretation    \\
\hline
$>5$        & Strongly disfavored / tensioned   \\
$2.5\sim5$  & Moderately disfavored / tensioned \\
$1\sim2.5$  & Weakly disfavored / tensioned     \\
$-1\sim1$   & Inconlusive                       \\
$-2.5\sim-1$& Weakly favored / aligned          \\
$-5\sim-2.5$& Moderately favored / aligned      \\
$<-5$       & Strongly favored / aligned        \\
\hline
\end{tabular}
\caption{\label{table:jeffreys}
Jeffreys' scale for deciding the evidence of model $M_1$ over $M_2$ or the tension between datasets.}
\end{table}

\section{Prior volume bias analysis}
\label{sec:prior}

\begin{table}[]
\centering
\begin{tabular}{|c||c|c|c|c|c|c|}
\hline
                    & \multicolumn{3}{c|}{$\phi$CDM}    & \multicolumn{3}{c|}{$\Lambda$CDM} \\
\hline
Parameters          & CMB       & + BAO     & + SNe     & CMB       & + BAO     & +SNe      \\
\hline
$\log(10^9 A_s)$    & $-0.16$   & $+0.11$   & $-0.05$   & $+0.19$   & $-0.04$   & $-0.32$   \\
$n_s$               & $+0.39$   & $+0.20$   & $-0.19$   & $-0.40$   & $+0.22$   & $+0.27$   \\
$\Omega_{b0}h^2$    & $+0.31$   & $+0.34$   & $+0.06$   & $-0.03$   & $+0.15$   &{\color{Orange}$-0.55$}\\
$\Omega_{c0}h^2$    & {\color{Red}$-0.70$}  & $-0.11$   & $+0.06$   & $+0.15$   & $\pm0.00$ & $-0.12$   \\
$\tau_\mathrm{reio}$& $-0.08$   & $+0.09$   & $-0.15$   & $+0.03$   & $+0.09$   & $-0.25$   \\
$100\theta_{MC}$    & $+0.12$   & $-0.33$   & {\color{Red}$+0.75$}  & $+0.35$   & {\color{Red}$+0.75$}  & $-0.13$   \\
\hline
$y_\mathrm{cal}$    & $+0.06$   & $+0.31$   & $+0.02$   & $+0.26$   & $-0.38$   & $-0.05$   \\
$A_{143}$           & $-0.31$   & $+0.32$   & $-0.13$   & $+0.32$   & $-0.03$   & $+0.22$   \\
$A_{217}$           & $-0.00$   & $+0.15$   & $-0.29$   & $+0.25$   & $+0.06$   & $+0.25$   \\
$A_{143\times217}$  & $-0.13$   & $+0.21$   & $-0.23$   & $+0.24$   & $-0.04$   & $+0.17$   \\
$\gamma_{143}$      & $+0.33$   & $+0.04$   & $+0.22$   & $-0.08$   & $-0.14$   & $-0.12$   \\
$\gamma_{217}$      & $-0.05$   & $-0.21$   & $+0.31$   & $-0.07$   & $-0.15$   & $-0.35$   \\
$\gamma_{143\times217}$ &$+0.02$& $-0.34$   & $+0.22$   & $-0.38$   & $-0.36$   & $-0.44$   \\
$c_{TE}$            & $-0.37$   & $+0.25$   & $+0.20$   & $+0.17$   & $+0.24$   &{\color{Orange}$+0.61$}\\
$c_{EE}$            & $-0.27$   & $-0.24$   & $+0.18$   & {\color{Red}$+0.69$}  & $-0.03$   & $+0.15$   \\
\hline
$H_0$               & {\color{Red}$+0.74$}  & $+0.32$   & $+0.15$   & $-0.07$   & $+0.20$   & $-0.09$   \\
$\Omega_{m0}$       & {\color{Red}$-0.76$}  & $-0.26$   & $-0.11$   & $+0.09$   & $-0.14$   & $-0.02$   \\
$S_8$               & {\color{Red}$-0.81$}  & $-0.04$   & $-0.04$   & $+0.18$   & $-0.08$   & $-0.19$   \\
\hline
$\mathcal{E}_{H_0}$ & $+0.03$   &{\color{Orange}$-0.58$}& {\color{Red}$-1.07$}  &&&\\
$n$                 & {\color{Red}$-0.69$}  & {\color{Red}$+0.79$}  & {\color{Red}$-0.87$}  &&&\\
$\log_{10}(\rho_\mathrm{DE,0}/V_\mathrm{min} - 1)$  & {\color{Red}$-0.89$}  & {\color{Red}$-1.54$}  & {\color{Red}$+1.29$}  &&&\\
$\log_{10}(n \eta^{-2})$    & {\color{Red}$+0.95$}  & {\color{Red}$+0.86$}  & {\color{Red}$-1.01$}  &&&\\
$\log_{10}(\phi_i / \eta)$  & {\color{Red}$+0.79$}  & $-0.10$   &{\color{Orange}$+0.58$}&&&\\
\hline
\end{tabular}
\caption{The shift of model and nuisance parameters between maximum-a-posteriori (MAP) value and Bayesian mean in unit of Bayesian standard deviation $( \theta_\mathrm{map} - \left\langle \theta \right\rangle )/ \sqrt{\left\langle \theta^2 \right\rangle - \left\langle \theta \right\rangle^2}$ for the axion-like dark energy model ($\phi$CDM) and $\Lambda$CDM model in CMB, CMB + BAO, and CMB + BAO + SNe dataset. Deviation between $0.45\sim0.65\sigma$ and above is marked {\color{Orange}orange} and {\color{Red}red} respectively.}
\label{table:MAP_vs_mean}
\end{table}

As made apparent in Figs.~\ref{fig:fit_scatter_SN_LowZ} and \ref{fig:fit_scatter_CMB}, the constraint on the extended parameters is rather weak. This makes the choice of the prior of these parameters in table~\ref{table:prior} the dominant factor in Bayesian analysis. Different priors, especially those biasing toward specific direction, could lead to Bayesian mean values different from the maximal a posteriori (MAP) value, signalling the breakdown of the Gaussian approximation. We verify that the Bayesian mean does not deviate far away from MAP value in the analysis of section~\ref{sec:fit} as shown in table~\ref{table:MAP_vs_mean} other than the extended parameters and the Gaussian constraint $\mathcal{E}_{H_0}$, with the sole exception of the axion-like dark energy model against the CMB dataset where the distribution of multiple observables does become non-Gaussian.

This is consistent with the conclusion drawn from Fig.~\ref{fig:fit_scatter_CMB} in section~\ref{sec:result}, that without any late-time measurement, the CMB dataset alone permits strong modification of the cosmic evolution, i.e., a dark energy with an equation of state deviating away from $-1$ during the dark-energy-dominated era. The finite prior volume of $n$ thus biases the posterior toward $\Lambda$CDM-like configurations ($n\sim 0$). This bias is further enhanced by the finite width of the Gaussian constraint that greatly expands the prior volume around $\log_{10} (\rho_{\rm DE,0} / V_{\rm min} - 1) \sim -3$ and $n\eta^{-2} \gg 1$.

We emphasise again that this biasing effect on the Bayesian analysis is fully anticipated, and is in fact the key feature of the axion-like dark energy model considered in this work, i.e. a mechanism for the grand entrance and graceful exit of the tracking regime. The sliding of the tracking regime guarantees a finite window within an infinite parameter space where the axion-like dark energy behaves differently than a cosmological constant. Therefore, the prior volume ratio between that finite window and the infinite $\Lambda$CDM-like region will always introduce bias into the analysis.

\begin{figure}
\centering
\includegraphics[width = 0.96 \textwidth]{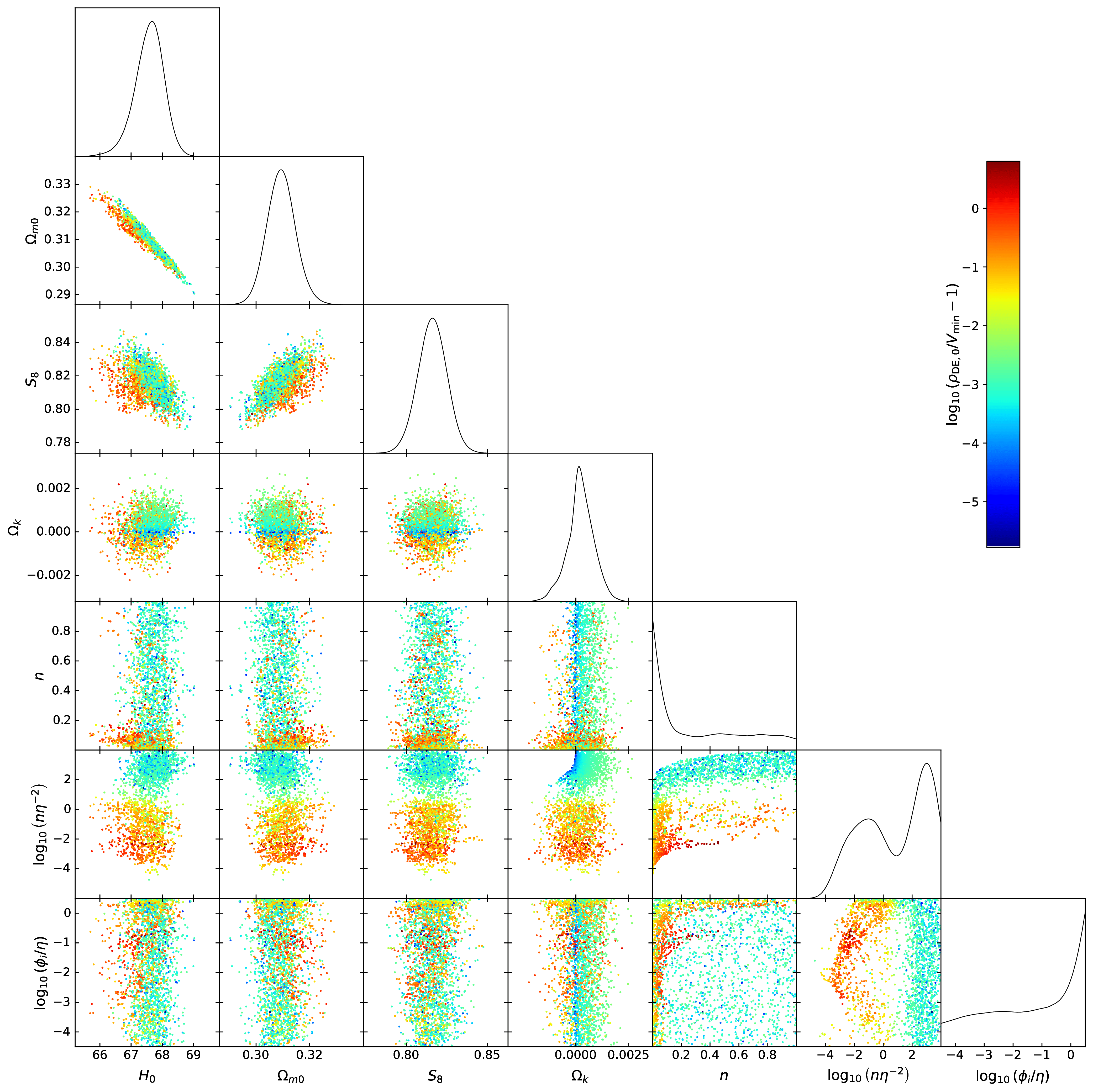}
\caption{\label{fig:fit_scatter_CMB_BAO_SN} Scatter plot of the late-time parameters, the extended parameters and the Gaussian constraint for the axion-like dark energy model against CMB + BAO + SNe dataset, colour-coded by $\log_{10} (\rho_{\rm DE,0} / V_{\rm min} - 1)$.
}
\end{figure}

\begin{figure}
\centering
\includegraphics[width = 0.96 \textwidth]{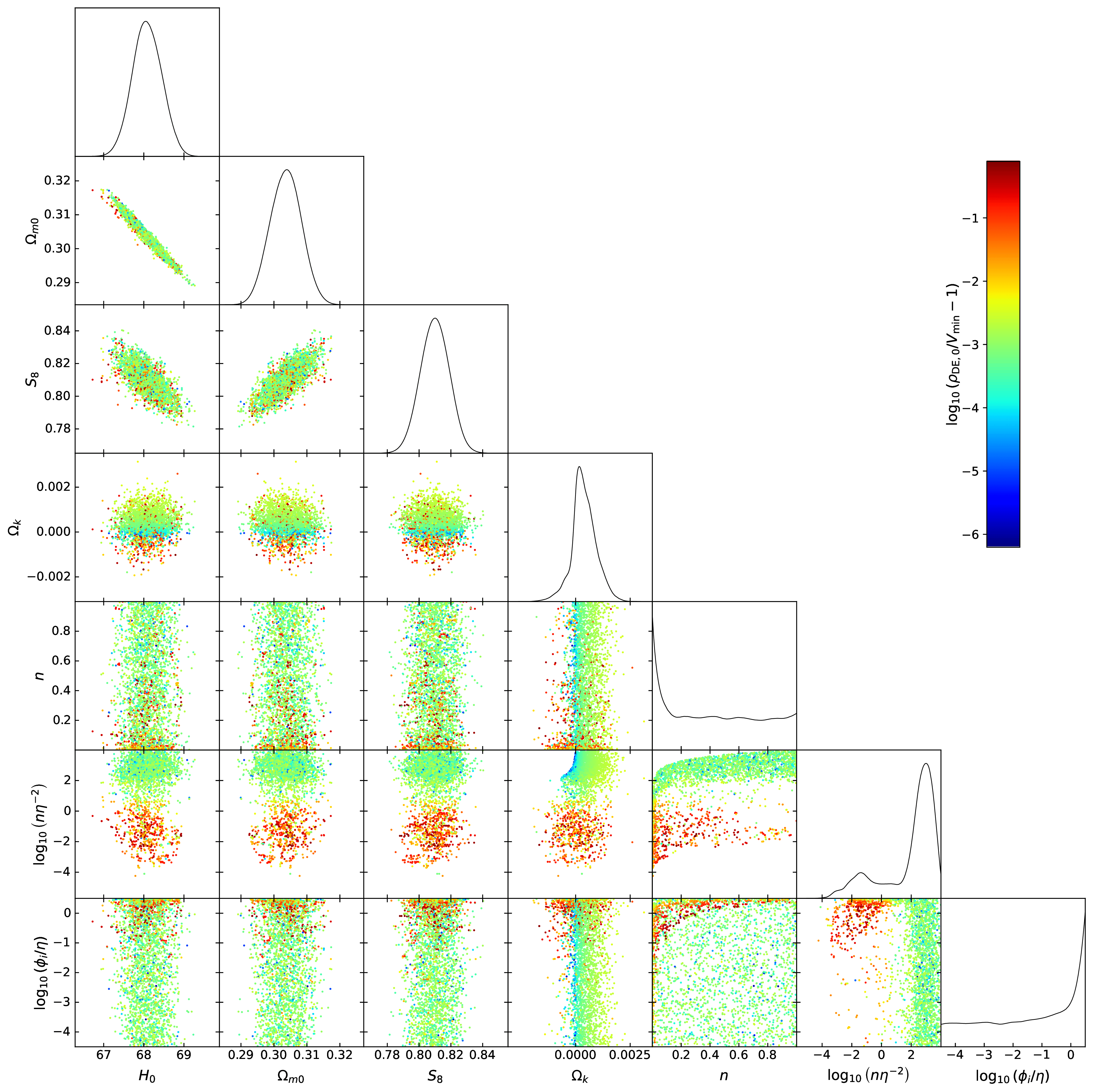}
\caption{\label{fig:fit_scatter_CMB_BAO_SN_LowZ} Scatter plot of the late-time parameters, the extended parameters and the Gaussian constraint for the axion-like dark energy model against CMB + BAO + SNe + low-$z$ dataset, colour-coded by $\log_{10} (\rho_{\rm DE,0} / V_{\rm min} - 1)$.
}
\end{figure}

As evidently shown in table~\ref{table:MAP_vs_mean}, such a bias toward $\Lambda$CDM is clearly suppressed once considering distance anchors, as the dark energy equation of state is constrained to be close to $-1$. However, the finite width effect of the Gaussian constraint remains, as it introduces bias toward $n\eta^{-2} \gg 1$ as show in Figs.~\ref{fig:fit_scatter_CMB_BAO_SN} and \ref{fig:fit_scatter_CMB_BAO_SN_LowZ}. This again leads to bias toward $\Lambda$CDM in the case of CMB + BAO + SNe dataset, but not in the case of CMB + BAO + SNe + low-$z$ dataset as the distance measurements are finally strong enough to exclude all configurations of the axion-like dark energy that behaves differently from $\Lambda$CDM in the dark-energy-dominated era.

\acknowledgments
HWC is supported by the NSFC Grant No.~12250410250 and No.~12347133. 
C. G. B. acknowledges financial support from the FPI fellowship PRE2021-100340 of the Spanish Ministry of Science, Innovation and Universities. M.B.-L. is grateful to the hospitality of LeCosPA, NTU (Taiwan) where this work was initiated during one of her visits. M. B.-L. is supported by the Basque Foundation of Science Ikerbasque. Our work is supported by the Spanish Grant PID2023-149016NB-I00 (MINECO/AEI/FEDER, UE). This work is also supported by the Basque government Grant No. IT1628-22 (Spain). 
We are also grateful to Jes{\'u}s Torrado for his assistance with the Cobaya software during the early stages of this work. This article
is based upon work from the COST Action CA21136
“Addressing observational tensions in cosmology with
systematics and fundamental physics” (CosmoVerse),
supported by COST (European Cooperation in Science
and Technology).









\bibliographystyle{JHEP}

\bibliography{bibliography.bib}

\end{document}